\documentclass{article}
\usepackage{amsmath}
\usepackage{amsfonts}

\newtheorem{theorem}{Theorem}

\newtheorem{lemma}[theorem]{Lemma}

\newtheorem{proposition}[theorem]{Proposition}

\input{tcilatex}

\begin{document}

\title{Isometry theorem for the Segal--Bargmann transform on a noncompact
symmetric space of the complex type}
\author{Brian C. Hall\thanks{%
Supported in part by NSF grants DMS-0200649 and DMS-0555862} and Jeffrey J.
Mitchell \\
bhall{@}nd.edu and mitchellj{@}rmu.edu}
\date{August 2007}
\maketitle

\begin{abstract}
We consider the Segal--Bargmann transform on a noncompact symmetric space of
the complex type. We establish isometry and surjectivity theorems for the
transform, in a form as parallel as possible to the results in the dual
compact case. The isometry theorem involves integration over a tube of
radius $R$ in the complexification, followed by analytic continuation with
respect to $R.$ A cancellation of singularities allows the relevant integral
to have a nonsingular extension to large $R,$ even though the function being
integrated has singularities.
\end{abstract}

\section{Introduction}

\subsection{Euclidean and compact cases}

The Segal--Bargmann transform for the Euclidean space $\mathbb{R}^{d},$ in a
form convenient for the present paper, can be expressed as follows. Let $t$
be a fixed positive number and let $e^{t\Delta /2}$ be the time-$t$ forward 
\textit{heat operator} for $\mathbb{R}^{d}$. It is not hard to show that for
any $f$ in $L^{2}(\mathbb{R}^{d},dx)$, $e^{t\Delta /2}f$ admits an entire
analytic continuation in the space variable from $\mathbb{R}^{d}$ to $%
\mathbb{C}^{d}.$ The Segal--Bargmann transform \cite{Ba1,Se3} is then the
map associating to each $f\in L^{2}(\mathbb{R}^{d})$ the holomorphic
function obtained by analytically continuing $e^{t\Delta /2}f$ from $\mathbb{%
R}^{d}$ to $\mathbb{C}^{d}.$ Basic properties of this transform are encoded
in the following theorem. (See \cite{range} for more information.)

\begin{theorem}
\label{rd.thm}The \textbf{isometry formula}. Fix $f$ in $L^{2}(\mathbb{R}%
^{d},dx).$ Then the function $F:=e^{t\Delta /2}f$ has an analytic
continuation to $\mathbb{C}^{d}$ satisfying%
\begin{equation}
\int_{\mathbb{R}^{d}}\left\vert f(x)\right\vert ^{2}~dx=\int_{\mathbb{C}%
^{d}}\left\vert F(x+iy)\right\vert ^{2}\frac{e^{-\left\vert y\right\vert
^{2}/t}}{(\pi t)^{d/2}}~dy~dx.  \label{rdisom}
\end{equation}

The \textbf{surjectivity theorem}. Given any holomorphic function $F$ on $%
\mathbb{C}^{d}$ for which the right-hand side of (\ref{rdisom}) is finite,
there exists a unique $f\in L^{2}(\mathbb{R}^{d})$ with $\left. F\right\vert
_{\mathbb{R}^{d}}=e^{t\Delta /2}f.$

The \textbf{inversion formula}. If $f\in L^{2}(\mathbb{R}^{d})$ is
sufficiently regular and $F:=e^{t\Delta /2}f,$ then%
\begin{equation*}
f(x)=\int_{\mathbb{R}^{d}}F(x+iy)\frac{e^{-\left\vert y\right\vert ^{2}/2t}}{%
(2\pi t)^{d/2}}~dy.
\end{equation*}
\end{theorem}

Note that we have $e^{-\left\vert y\right\vert ^{2}/t}$ in the isometry
formula but $e^{-\left\vert y\right\vert ^{2}/2t}$ in the inversion formula.
From the point of view of harmonic analysis, the Segal--Bargmann transform
may be thought of as a way of combining information about a function $f(x)$
on $\mathbb{R}^{d}$ with information about the Fourier transform $\hat{f}(y)$
of $f$ into a single (holomorphic) function $F(x+iy)$ on $\mathbb{C}^{d}=%
\mathbb{R}^{2d}.$ From the point of view of quantum mechanics, $F$ may be
thought of as the phase space wave function corresponding to the position
space wave function $f.$ For more information, see \cite{Ba1, mexnotes,
lpbounds,Fo}.

Analogous results for compact symmetric spaces have been obtained by Hall 
\cite{H1,H2} in the compact group case and by Stenzel \cite{St1} in the
general case. (See \cite{bull, range} for more information. See also \cite%
{KTX} for surprising results in the case of the Heisenberg group.) Let $X$
denote a compact symmetric space, assumed for simplicity to be simply
connected. Then $X$ can be expressed as $X=U/K,$ where $U$ is a simply
connected compact Lie group and $K$ is the fixed-point subgroup of an
involution. We may define the complexification of $U/K$ to be $U_{\mathbb{C}%
}/K_{\mathbb{C}},$ where $U_{\mathbb{C}}$ is the unique simply connected Lie
group whose Lie algebra is $\mathfrak{u}+i\mathfrak{u}$ and where $K_{%
\mathbb{C}}$ is the connected Lie subgroup of $U_{\mathbb{C}}$ whose Lie
algebra is $\mathfrak{k}+i\mathfrak{k}.$ Then $U_{\mathbb{C}}/K_{\mathbb{C}}$
may be identified diffeomorphically with the tangent bundle $T(U/K)$ by
means of the map $\Phi :T(U/K)\rightarrow U_{\mathbb{C}}/K_{\mathbb{C}}$
given by%
\begin{equation}
\Phi (x,Y)\rightarrow \exp _{x}(iY),  \label{compact.ident}
\end{equation}%
where $Y$ is a tangent vector to $U/K$ at $x$ and where $\exp _{x}(iY)$
refers to the analytic continuation of the geometric exponential map for $%
U/K.$ See \cite[Eq. 2]{HM1} for a simple explicit formula for $\Phi (x,Y)$
in the case that $U/K$ is a sphere.

If the Lie algebra $\mathfrak{u}$ of $U$ is decomposed in the usual way as $%
\mathfrak{u}=\mathfrak{k}+\mathfrak{p},$ then let $G$ be the connected Lie
subgroup of $U_{\mathbb{C}}$ whose Lie algebra is $\mathfrak{g}=\mathfrak{k}%
+i\mathfrak{p}.$ The \textit{dual noncompact symmetric space} to $U/K$ is
the manifold $G/K$, equipped with an appropriate $G$-invariant Riemannian
metric. The identification (\ref{compact.ident}) of $T(U/K)$ with $U_{%
\mathbb{C}}/K_{\mathbb{C}}$ gives rise to an identification of each fiber in 
$T(U/K)$ with $G/K.$ Specifically, if $x_{0}$ is the identity coset in $U/K,$
then the image of $T_{x_{0}}(U/K)$ under $\Phi $ is precisely the $G$-orbit
of the identity coset in $U_{\mathbb{C}}/K_{\mathbb{C}}.$ Furthermore, the
stabilizer in $G$ of the identity coset is precisely $K,$ and so $\Phi
(T_{x_{0}}(U/K))\cong G/K.$ Any other fiber in $T(U/K)$ is then identified
with $T_{x_{0}}(U/K)\cong G/K$ by the action of $U.$ See \cite{St1, range}
for details.

Having identified each tangent space $T_{x}(U/K)$ with the noncompact
symmetric space $G/K,$ we have on each tangent space the \textit{heat kernel
density }$\nu _{t}^{\mathrm{nc}}$ (based at the origin) and the \textit{%
Jacobian }$j^{\mathrm{nc}}$ \textit{of the exponential map} with respect to
the Riemannian metric for $G/K.$ Here the superscript \textquotedblleft $%
\mathrm{nc}$\textquotedblright\ indicates a quantity associated to the
noncompact symmetric space $G/K$ dual to the original compact symmetric
space $U/K.$ The result is then the following. (See \cite{St1}; compare \cite%
{H1, H2} in the compact group case.)

\begin{theorem}
\label{compact.thm}The \textbf{isometry formula}. Fix $f$ in $L^{2}(U/K)$.
Then the function $F:=e^{t\Delta /2}f$ has an analytic continuation to $U_{%
\mathbb{C}}/K_{\mathbb{C}}$ satisfying%
\begin{equation}
\int_{U/K}\left\vert f(x)\right\vert ^{2}~dx=\int_{x\in U/K}\int_{Y\in
T_{x}(U/K)}\left\vert F(\exp _{x}(iY))\right\vert ^{2}\nu _{2t}^{\mathrm{nc}%
}(2Y)j^{\mathrm{nc}}(2Y)~2^{d}dY~dx.  \label{compact.isom}
\end{equation}%
Here $d=\dim (U/K),$ $dY$ is the Lebesgue measure on $T_{x}(U/K),$ and $dx$
is the Riemannian volume measure on $U/K.$

The \textbf{surjectivity theorem}. Given any holomorphic function $F$ on $U_{%
\mathbb{C}}/K_{\mathbb{C}}$ for which the right-hand side of (\ref%
{compact.isom}) is finite, there exists a unique $f\in L^{2}(U/K)$ with $%
\left. F\right\vert _{U/K}=e^{t\Delta /2}f.$

The \textbf{inversion formula}. If $f\in L^{2}(U/K)$ is sufficiently regular
and $F:=e^{t\Delta /2}f,$ then%
\begin{equation}
f(x)=\int_{T_{x}(U/K)}F(\exp _{x}(iY))\nu _{t}^{\mathrm{nc}}(Y)j^{\mathrm{nc}%
}(Y)~dY.  \label{compact.inv}
\end{equation}
\end{theorem}

Note that in the inversion formula we have $\nu _{t}(Y)j(Y)$ whereas in the
isometry formula we have $\nu _{2t}(2Y)j(2Y).$ Note also that the isometry
and inversion formulas for Euclidean space are of the same form as Theorem %
\ref{compact.thm}, with $\exp _{x}(iy)=x+iy,$ $j(y)\equiv 1,$ and $\nu
_{t}(y)=(2\pi t)^{-d/2}e^{-\left\vert y\right\vert ^{2}/2t}.$

An important special case of Theorem \ref{compact.thm} is the compact group
case considered in \cite{H1, H2}, i.e., the case in which $K$ is the
diagonal subgroup of $U=K\times K.$ This case is connected to stochastic
analysis and the Gross ergodicity theorem \cite{GM,HS,ergodic} and to the
quantization of Yang--Mills theory on a spacetime cylinder \cite{Wr,DH1,
ymcoherent}. Furthermore, in this case the isometry formula can be
understood as a unitary pairing map in the context of geometric quantization 
\cite{geoquant, FMMN1, FMMN2,Ty}.

In the compact group case, the dual noncompact symmetric space is of the
\textquotedblleft complex type,\textquotedblright\ and in this case there is
a simple explicit formula for the heat kernel $\nu _{t}^{\mathrm{nc}},$
namely,%
\begin{equation}
\nu _{t}^{\mathrm{nc}}(Y)=e^{-\left\vert \rho \right\vert ^{2}t/2}j^{\mathrm{%
nc}}(Y)^{-1/2}\frac{e^{-\left\vert Y\right\vert ^{2}/2t}}{(2\pi t)^{d/2}}.
\label{compact.nut}
\end{equation}%
Here $\rho $ is half the sum (with multiplicity) of the positive roots for $%
G/K$ and there is a simple explicit expression for $j^{\mathrm{nc}}$ (change 
$\sin $ to $\sinh $ in (\ref{jc})). Thus, in the compact group case, the
isometry formula takes the form%
\begin{equation}
\int_{U/K}\left\vert f(x)\right\vert ^{2}~dx=e^{-\left\vert \rho \right\vert
^{2}t}\int_{x\in U/K}\int_{Y\in T_{x}(U/K)}\left\vert F(\exp
_{x}(iY))\right\vert ^{2}j^{\mathrm{nc}}(2Y)^{1/2}\frac{e^{-\left\vert
Y\right\vert ^{2}/t}}{(\pi t)^{d/2}}~dY~dx  \label{group.isom}
\end{equation}%
and the inversion formula takes the form%
\begin{equation}
f(x)=e^{-\left\vert \rho \right\vert ^{2}t/2}\int_{T_{x}(U/K)}F(\exp
_{x}(iY))j^{\mathrm{nc}}(Y)^{1/2}\frac{e^{-\left\vert Y\right\vert ^{2}/2t}}{%
(2\pi t)^{d/2}}~dY.  \label{group.inv}
\end{equation}

\subsection{The complex case}

Since we have nice theories for the Euclidean and compact cases, the natural
next step is to consider symmetric spaces of the noncompact type. This would
mean applying the heat operator to a function on a symmetric space of the
form $G/K,$ where $G$ is a noncompact semisimple Lie group (connected with
finite center) and $K$ is a maximal compact subgroup. If we attempt to
imitate the constructions in the compact and Euclidean cases, we rapidly
encounter difficulties. As in the compact case, we can define a smooth map $%
\Phi :T(G/K)\rightarrow G_{\mathbb{C}}/K_{\mathbb{C}}$ by%
\begin{equation*}
\Phi (x,Y)=\exp _{x}(iY).
\end{equation*}%
However, in the noncompact case, $\Phi $ is \textit{not} a global
diffeomorphism; $\Phi $ is not globally injective and the differential of $%
\Phi $ becomes degenerate at certain points. The map $\Phi $ gives rise to a 
\textit{local} identification of each fiber in $T(G/K)$ with the dual 
\textit{compact} symmetric space, but this identification cannot possibly be
global, since $T_{x}(G/K)$ is not compact. In addition to the (global)
breakdown of the desired identifications, we have a problem with analytic
continuation. For a typical function $f$ in $L^{2}(G/K),$ the function $%
e^{t\Delta /2}f$ does not have a global analytic continuation to $G_{\mathbb{%
C}}/K_{\mathbb{C}},$ but rather becomes both singular and multiple valued
once one moves far enough from $G/K.$

The paper \cite{HM3} takes a first step in overcoming these obstacles.
(Related but nonoverlapping results were obtained by Kr\"{o}tz, \'{O}%
lafsson, and Stanton \cite{KOS}. We discuss \cite{KOS} in detail in Section %
\ref{kos.sub} and Section \ref{kos.sec}. See also \cite{OO,DOZ1,DOZ2} for a
different approach, not involving the heat equation.) In \cite{HM3}, we
consider the simplest case, that of noncompact symmetric spaces of the
\textquotedblleft complex type.\textquotedblright\ Here complex type does 
\textit{not} mean that the symmetric space is a complex manifold, but rather
that the group $G$ admits a complex structure, which means that $G$ is the
complexification of $K.$ The complex\ case is nothing but the noncompact
dual of the compact group case. The simplest symmetric space of the complex
type is hyperbolic 3-space, where $G\cong SO(3,1)_{e}\cong PSL(2,\mathbb{C}%
). $

In the complex case, we develop in \cite{HM3} (1) an isometry formula for
\textquotedblleft radial\textquotedblright\ (i.e., left-$K$-invariant)\
functions on $G/K$ and (2) an inversion formula for general functions
(sufficiently regular but not necessarily radial). Suppose $f$ is a radial
function in $L^{2}(G/K)$ and let $F=e^{t\Delta /2}f.$ Then the isometry
formula of \cite[Thm. 2]{HM3} states that the map $Y\rightarrow F(\exp
_{x_{0}}Y)$ has a meromorphic extension to $\mathfrak{p}_{\mathbb{C}}$ and
that the $L^{2}$ norm of $F$ over $\mathfrak{p}_{\mathbb{C}}$ with respect
to a certain measure $\mu $ is equal to the $L^{2}$ norm of $f$ over $G/K.$
See also \cite[Thm. 2.8]{OS}.

The inversion formula of \cite{HM3}, meanwhile, reads 
\begin{equation}
f(x)=\text{\textquotedblleft }\lim_{R\rightarrow \infty }\text{%
\textquotedblright }\ e^{\left\vert \rho \right\vert ^{2}t/2}\int_{\substack{
Y\in T_{x}(G/K)  \\ \left\vert Y\right\vert \leq R}}F(\exp _{x}iY)j^{\mathrm{%
c}}(Y)^{1/2}\frac{e^{-\left\vert Y\right\vert ^{2}/2t}}{(2\pi t)^{d/2}}~dY.
\label{complex.inv1}
\end{equation}%
(See \cite[Thm. 4]{HM3}. A different approach to inversion formulas is taken
in \cite{St2}.) Here $j^{\mathrm{c}}$ is the Jacobian of the exponential
mapping for the \textit{compact} symmetric space $U/K$ dual to $G/K$ and $%
c=\left\vert \rho \right\vert ^{2},$ where $\rho $ is half the sum (with
multiplicities) of the positive roots for $G/K$. Moreover, \textquotedblleft 
$\lim_{R\rightarrow \infty }${}\textquotedblright\ means that the integral
on the right-hand side of (\ref{complex.inv1}) is well-defined for all
sufficiently small $R$ and admits a real-analytic continuation in $R$ to $%
(0,\infty ).$ The right-hand side of (\ref{complex.inv1}) then is equal to
the limit as $R$ tends to infinity of this analytic continuation. That is, a
limit with quotation marks means the limit as $R$ tends to infinity of 
\textit{the real-analytic extension of} the indicated quantity.

It should be noted that although $F(\exp _{x}iY)$ develops singularities
once $Y$ gets sufficiently large, the integral on the right-hand side of (%
\ref{complex.inv1}) does not develop singularities; it has a real-analytic
extension to $R\in (0,\infty ).$ There is a delicate \textquotedblleft
cancellation of singularities\textquotedblright\ going on here, which is
explained in \cite{HM3}, \cite{range}, and the next subsection.

Leaving aside the analytic continuation in $R$, which is unnecessary in the
compact case, (\ref{complex.inv1}) is \textquotedblleft
dual\textquotedblright\ to the inversion formula (\ref{group.inv}) for the
compact group case. That is, (\ref{complex.inv1}) is obtained from (\ref%
{group.inv}) by changing $j^{\mathrm{nc}}$ to $j^{\mathrm{c}}$ and changing $%
e^{-\left\vert \rho \right\vert ^{2}t/2}$ to $e^{\left\vert \rho \right\vert
^{2}t/2}.$ (The constant $\left\vert \rho \right\vert ^{2}$ is related to
the scalar curvature, which is positive in the compact case and negative in
the noncompact case.)

The main result of the present paper is an isometry formula which bears the
same relationship to the inversion formula (\ref{complex.inv1}) as (\ref%
{group.isom}) bears to (\ref{group.inv}).

\begin{theorem}
\label{introisom.thm}For any $f$ in $L^{2}(G/K)$ ($G$ complex) we have%
\begin{eqnarray}
&&\int_{G/K}\left\vert f(x)\right\vert ^{2}~dx  \notag \\
&=&\text{\textquotedblleft }\lim_{R\rightarrow \infty }\text{%
\textquotedblright }~e^{\left\vert \rho \right\vert ^{2}t}\int_{x\in
G/K}\int _{\substack{ Y\in T_{x}(G/K)  \\ \left\vert Y\right\vert \leq R}}%
\left\vert F(\exp _{x}iY)\right\vert ^{2}j^{\mathrm{c}}(2Y)^{1/2}\frac{%
e^{-\left\vert Y\right\vert ^{2}/t}}{(\pi t)^{d/2}}~dY~dx.
\label{complex.isom1}
\end{eqnarray}
\end{theorem}

As in the inversion formula, the integral on the right-hand side of (\ref%
{complex.isom1}) is to be taken literally for small $R$ and interpreted by
means of analytic continuation in $R$ for large $R.$ See Theorem \ref%
{globalisom.thm} in Section \ref{globalisom.sec} for a more precise
statement. We will also prove a surjectivity theorem (Theorem \ref%
{surjectivity.thm} in Section \ref{surjectivity.sec}); roughly, if $F$ is
any holomorphic function on a $G$-invariant neighborhood of $G/K$ inside $G_{%
\mathbb{C}}/K_{\mathbb{C}}$ for which the right-hand side of (\ref%
{complex.isom1}) makes sense and is finite, then there exists a unique $f\in
L^{2}(G/K)$ with $\left. F\right\vert _{G/K}=e^{t\Delta /2}f.$

In the case of hyperbolic 3-space, with the usual normalization of the
metric, the isometry formula takes the following explicit form (see also 
\cite[Sect. 5]{range}):%
\begin{equation}
\int_{H^{3}}\left\vert f(x)\right\vert ^{2}~dx=\text{\textquotedblleft }%
\lim_{R\rightarrow \infty }\text{\textquotedblright }~e^{t}\int_{x\in
H^{3}}\int_{\substack{ Y\in T_{x}(H^{3})  \\ \left\vert Y\right\vert \leq R}}%
\left\vert F(\exp _{x}iY)\right\vert ^{2}\frac{\sin \left\vert 2Y\right\vert 
}{\left\vert 2Y\right\vert }\frac{e^{-\left\vert Y\right\vert ^{2}/t}}{(\pi
t)^{3/2}}~dY~dx.  \label{h3.isom}
\end{equation}

The isometry formula of Kr\"{o}tz, \'{O}lafsson, and Stanton \cite{KOS},
when specialized to the complex case, is \textit{not} the same as the
formula in \ref{introisom.thm}. We discuss the relationship between the two
results in Section \ref{kos.sub} and in Section \ref{kos.sec}. If $f$ just
happens to be radial, then there is another isometry formula, established in 
\cite[Thm. 2]{HM3} (see also \cite[Thm. 2.8]{OS}). For radial functions, it
is not immediately obvious how to see directly that the isometry formula in
Theorem \ref{introisom.thm} agrees with the isometry formula of \cite{HM3}.

\subsection{Cancellation of singularities}

Let $T^{R}(G/K)$ denote the set of $(x,Y)\in T(G/K)$ with $\left\vert
Y\right\vert <R.$ The inversion and isometry formulas assert that for
noncompact symmetric spaces of the complex type, certain integrals (those on
the right-hand side of (\ref{complex.inv1}) and (\ref{complex.isom1}))
involving $F(\exp _{x}iY)$ over $T^{R}(G/K)$ are \textquotedblleft
nonsingular,\textquotedblright\ in that they extend in a real analytic way
to all positive values of $R.$ On the other hand, $F(\exp _{x}iY)$ itself
does in fact become singular (and multiple-valued) once $Y$ gets
sufficiently large, as can be seen, for example, from the formula \cite[%
Prop. 3.2]{Ga} for the heat kernel on $G/K$. If $F(\exp _{x}iY)$ itself
becomes singular for large $Y$ but certain integrals involving $F$ remain
nonsingular, then some \textquotedblleft cancellation of
singularities\textquotedblright\ must be taking place in the process of
integration.

In the case of the inversion formula, the cancellation of singularities
occurs because the integral on the right-hand side of (\ref{complex.inv1})
only \textquotedblleft sees\textquotedblright\ the part of $F(\exp _{x}iY)$
that is \textquotedblleft radial\textquotedblright\ in $Y$ (i.e., invariant
under the adjoint action of $K$). Meanwhile, the radial part of $F(\exp
_{x}iY)$ can be expanded in terms of analytically continued spherical
functions. In the complex case, the analytically continued spherical
functions have only a very simple sort of singularity, a certain denominator
function (the same one for all spherical functions) that can become zero.
(See Section \ref{kos.sec} for precise formulas.) The zeros of this
denominator function are canceled by the zeros of the function $j^{\mathrm{c}%
}$ in the integrand of (\ref{complex.inv1}).

Meanwhile, in the isometry formula, the integral of $\left\vert F\right\vert
^{2}$ over $T^{R}(G/K)$ can be expressed as an integral of $\left\vert
F\right\vert ^{2}$ over $G$-orbits, followed by an integration over the
space of $G$-orbits in $T^{R}(G/K).$ Meanwhile, the Gutzmer-type formula of
Faraut \cite{Far1,Far2} (also used in an important way in \cite{KOS}) shows
that the orbital integrals of $\left\vert F\right\vert ^{2}$ can again be
expressed in terms of the analytically continued spherical functions. As in
the case of the inversion formula, the singularities coming from the
analytically continued spherical functions are (in the complex case)
canceled by the zeros of $j^{\mathrm{c}}$ in the integrand in (\ref%
{complex.isom1}). See (\ref{alphaint.complex}) and the discussion following
it. In the $H^{3}$ case, the integral of $|F(\exp _{x}iY)|^{2}$ over the set
of $(x,Y)$ with $\left\vert Y\right\vert =R$ blows up at $R=\pi /2$ like $%
1/\sin 2R.$ This blow-up is canceled by the factor of $\sin \left\vert
2Y\right\vert $ in (\ref{h3.isom}).

From a more philosophical point of view, we note work of R. Sz\H{o}ke \cite%
{Sz2}. Sz\H{o}ke has shown that although the differential of the map $\Phi
:T(G/K)\rightarrow G_{\mathbb{C}}/K_{\mathbb{C}}$ becomes degenerate at
certain points, the pullback of the $(1,0)$ sub-bundle of $T_{\mathbb{C}}(G_{%
\mathbb{C}}/K_{\mathbb{C}})$ by means of $\Phi $ has a real-analytic
extension to the whole of $T(X).$ The problem is that this bundle has
nonzero intersection with its complex-conjugate at certain points.
Nevertheless, Sz\H{o}ke's result suggests that things do not break down
entirely when the differential of $\Phi $ becomes degenerate.

\subsection{The results of Kr\"{o}tz, \'{O}lafsson, and Stanton\label%
{kos.sub}}

We now give a quick comparison of our isometry formula to the one of B. Kr%
\"{o}tz, G. \'{O}lafsson, and R. Stanton established in \cite{KOS}; details
are provided in Section \ref{kos.sec}. The paper \cite{KOS} establishes an
isometry formula for the Segal--Bargmann transform on an arbitrary globally
symmetric space $G/K$ of the noncompact type, with $G$ not necessarily
complex. The authors of \cite{KOS} consider the integral of $\left\vert
F\right\vert ^{2}$ over $G$-orbits in a certain open subset $\Xi $ of $G_{%
\mathbb{C}}/K_{\mathbb{C}}$. These $G$-orbits are parameterized by points in
a certain open subset $2i\Omega $ of $i\mathfrak{a},$ where $\mathfrak{a}$
is a maximal commutative subspace of $\mathfrak{p}.$ Thus, we obtain the
orbital integral $\mathcal{O}_{\left\vert F\right\vert ^{2}}(iY),$ denoting
the integral of $\left\vert F\right\vert ^{2}$ over the $G$-orbit
parameterized by $iY\in 2i\Omega \subset i\mathfrak{a}.$ Kr\"{o}tz, \'{O}%
lafsson, and Stanton then show that there is a certain \textquotedblleft
shift operator\textquotedblright\ $D$ such that $D\mathcal{O}_{\left\vert
F\right\vert ^{2}}$ has a real-analytic extension from $2i\Omega $ to all of 
$i\mathfrak{a}.$ The isometry formula, Theorem 3.3 of \cite{KOS}, then
asserts that $\int_{G/K}\left\vert f(x)\right\vert ^{2}dx$ is equal to the
integral of $D\mathcal{O}_{\left\vert F\right\vert ^{2}}$ over $i\mathfrak{a}
$ with respect to a certain Gaussian measure.

In the complex case, the isometry formula of \cite{KOS} \textit{does not}
coincide with the one we establish in this paper. Nevertheless, the two
isometry formulas are equivalent in a sense that we explain in Section \ref%
{kos.sec}. Specifically, in the complex case, $D$ is a differential operator
and we will show that an integration by parts can turn the isometry formula
of \cite{KOS}\ into the one we prove here. (See also the recent preprint 
\cite{OS2}, which gives a another description of the image of the
Segal--Bargmann, different from both \cite{KOS} and the present paper.)

In the complex case, the form of the isometry formula in (\ref{complex.isom1}%
) seems preferable to the form in \cite{KOS}, simply because (\ref%
{complex.isom1}) is more parallel to what one has in the dual compact case (%
\ref{group.isom}). On the other hand, the result of \cite{KOS} is more
general, because it holds for arbitrary symmetric spaces of the noncompact
type, not just the complex case. It would be desirable to attempt to carry
out this integration by parts in general (not just in the complex case), so
as to recast the isometry formula of \cite{KOS} into a form more parallel to
what one has in the general compact case in (\ref{compact.isom}). However,
because the singularities in the orbital integral are more complicated once
one moves away from the complex case, it remains to be seen whether this
integration by parts can be carried out in general.

\section{Preliminaries}

Although our main result holds only for the complex case, it is instructive
to begin in the setting of a general symmetric space of the noncompact type
and then specialize when necessary to the complex case. We consider, then, a
connected semisimple Lie group $G$ with finite center, together with a fixed
maximal compact subgroup $K$ of $G.$ For our purposes, there is no harm in
assuming that $G$ is contained in a simply connected complexification $G_{%
\mathbb{C}}.$ There is a unique involution of $G$ whose fixed points are $K,$
and this leads to a decomposition of the Lie algebra $\mathfrak{g}$ of $G$
as $\mathfrak{g}=\mathfrak{k}+\mathfrak{p},$ where $\mathfrak{p}$ is the
subspace of $\mathfrak{g}$ on which the associated Lie algebra involution
acts as $-I.$ The spaces $\mathfrak{k}$ and $\mathfrak{p}$ satisfy $[%
\mathfrak{k},\mathfrak{p}]\subset \mathfrak{p}$ and $[\mathfrak{p},\mathfrak{%
p}]\subset \mathfrak{k}.$

We choose on $\mathfrak{p}$ an inner product invariant under the adjoint
action of $K.$ We then consider the manifold $G/K$ and we let $x_{0}$ denote
the identity coset in $G/K.$ We identify the tangent space to $G/K$ at $%
x_{0} $ with $\mathfrak{p}.$ The choice of an Ad-$K$-invariant inner product
on $\mathfrak{p}$ gives rise to a Riemannian metric on $G/K$ that is
invariant under the left action of $G.$ The manifold $G/K$, together with a
metric of this form, is a symmetric space of the noncompact type, in the
terminology of \cite{He1}.

In the Lie algebra $\mathfrak{g}_{\mathbb{C}}$ of $G_{\mathbb{C}},$ we
consider the subalgebra $\mathfrak{u}:=\mathfrak{k}+i\mathfrak{p}.$ We let $%
U $ denote the connected Lie subgroup of $G_{\mathbb{C}}$ whose Lie algebra
is $\mathfrak{u}.$ The inner product on $\mathfrak{p}$ induces an inner
product on $i\mathfrak{p}$ in an obvious way. This inner product determines
a Riemannian metric on $U/K$ invariant under the left action of $U$, and $%
U/K $ with this metric is a Riemannian symmetric space of the compact type,
known as the \textquotedblleft compact dual\textquotedblright\ of $G/K.$

Let $\mathfrak{a}$ be any maximal commutative subspace of $\mathfrak{p}.$
Let $\Sigma \subset \mathfrak{a}$ denote the set of (restricted) roots for
the pair $(\mathfrak{g},\mathfrak{k})$, where we use the inner product on $%
\mathfrak{p}$, restricted to $\mathfrak{a},$ to identify $\mathfrak{a}$ with 
$\mathfrak{a}^{\ast }.$ Let $\Sigma ^{+}$ denote a set of positive roots.
Let $W$ denote the Weyl group, that is, the subgroup of the orthogonal group
of $\mathfrak{a}$ generated by the reflections associated to $\alpha \in R.$
It is known that any vector in $\mathfrak{p}$ can be moved into $\mathfrak{a}
$ by the adjoint action of $K,$ and that the resulting vector in $\mathfrak{a%
}$ is unique up to the action of $W.$ We let $\mathfrak{a}^{+}$ denote the
closed fundamental Weyl chamber, that is, the set of points $Y$ in $%
\mathfrak{a}$ with $\alpha (Y)\geq 0$ for all $\alpha \in R^{+}.$ Then each
Weyl-group orbit contains exactly one point in $\mathfrak{a}^{+}.$

We let $\Omega $ denote the Weyl-invariant domain in $\mathfrak{a}$ given by%
\begin{equation}
\Omega =\left\{ Y\in \mathfrak{a}\left\vert \left\vert \left\langle \alpha
,Y\right\rangle \right\vert <\frac{\pi }{2}\right. \text{ for all }\alpha
\in \Sigma \right\} .  \label{omega.def}
\end{equation}%
We may think of $\Omega $ as a subset of $\mathfrak{p}=T_{x_{0}}(G/K).$ We
then define a set $\Lambda $ by%
\begin{equation}
\Lambda =G\cdot \Omega \subset T(G/K);  \label{lambda.def}
\end{equation}%
that is, $\Lambda $ is the smallest $G$-invariant set in $T(G/K)$ containing 
$\Omega .$ Thus, to determine if a point $Y\in T_{x}(G/K)$ belongs to $%
\Lambda ,$ we move $Y$ to a vector $Y^{\prime }\in T_{x_{0}}(G/K)$ by the
action of $G$ and then move $Y^{\prime }$ to a vector $Y^{\prime \prime }\in 
\mathfrak{a}$ by the action of $K.$ Then $Y\in \Lambda $ if and only if $%
Y^{\prime \prime }\in \Omega .$

We now consider a map $\Phi :T(G/K)\rightarrow G_{\mathbb{C}}/K_{\mathbb{C}}$
given by%
\begin{equation}
\Phi (x,Y)=\exp _{x}(iY),\quad x\in G/K,~Y\in T_{x}(G/K).  \label{phi.map}
\end{equation}%
Explicitly, we may identify $T(G/K)$ with $(G\times \mathfrak{p})/K,$ where $%
K$ acts on $G$ by right-multiplication and on $\mathfrak{p}$ by $%
Y\rightarrow k^{-1}Yk.$ In that case, the geometric exponential map from $%
T(G/K)$ to $G/K$ is given by $(g,Y)\rightarrow ge^{Y}K_{\mathbb{C}}$ and so $%
\Phi $ may also be expressed as%
\begin{equation}
\Phi (g,Y)=ge^{iY}K_{\mathbb{C}},\quad g\in G,~Y\in \mathfrak{p}.
\label{exp}
\end{equation}%
Here we observe that for $k\in K,$ $\Phi (gk,k^{-1}Yk)=\Phi (g,Y),$ so that $%
\Phi ,$ written as a map of $G\times \mathfrak{p}$ into $G_{\mathbb{C}}/K_{%
\mathbb{C}}$ descends to a map of $(G\times \mathfrak{p})/K$ into $G_{%
\mathbb{C}}/K_{\mathbb{C}}.$ From (\ref{exp}) we can see that $\Phi $ is a
globally defined smooth map of $T(G/K)$ into $G_{\mathbb{C}}/K_{\mathbb{C}}.$

In contrast to the analogous map in the compact case, $\Phi $ is not a
diffeomorphism of $T(G/K)$ onto $G_{\mathbb{C}}/K_{\mathbb{C}}.$ Indeed, $%
\Phi $ is not globally injective and $\Phi $ is not even a local
diffeomorphism near certain points in $T(G/K).$ Nevertheless, $\Phi $ maps $%
\Lambda $ diffeomorphically onto its image in $G_{\mathbb{C}}/K_{\mathbb{C}%
}. $ This image, denoted $\Xi $ in \cite{KOS}, is the Akhiezer--Gindikin
\textquotedblleft crown domain\textquotedblright\ \cite{AG}.\ That is,%
\begin{equation}
\Xi =\left\{ \left. \exp _{x}(iY)\in G_{\mathbb{C}}/K_{\mathbb{C}%
}\right\vert (x,Y)\in \Lambda \right\} .  \label{xi.def}
\end{equation}%
We will consistently parameterize points $z\in \Xi $ as $z=\exp _{x}(iY)$
with $(x,Y)\in \Lambda .$ We let 
\begin{equation*}
T^{R}(G/K)=\left\{ \left. (x,Y)\right\vert ~\left\vert Y\right\vert
<R\right\} .
\end{equation*}%
Then $T^{R}(G/K)$ is contained in $\Lambda $ for all sufficiently small $R.$
We let $R_{\max }$ denote the largest $R$ with this property:%
\begin{equation}
R_{\max }=\max \left\{ R\left\vert T^{R}(G/K)\subset \Lambda \right.
\right\} .  \label{rmax}
\end{equation}

The complex structure on $\Xi $ (as an open subset of the complex manifold $%
G_{\mathbb{C}}/K_{\mathbb{C}}$) can be transferred to $\Lambda $ by the
diffeomorphism $\Phi .$ This complex structure on $\Lambda \subset T(G/K)$
is in fact the \textquotedblleft adapted complex
structure\textquotedblright\ developed in \cite{GStenz1,GStenz2,LS,Sz1}.
Indeed, $\Lambda $ is the maximal connected domain in $T(G/K)$ containing
the zero section on which the adapted complex structure is defined. See \cite%
{AG,BHH,KS1,KS2} for more information.

\section{Partial isometry for general symmetric spaces of the noncompact
type \label{partial_gen.sec}}

We continue to work on $G/K,$ with $G$ arbitrary real semisimple (connected
with finite center), not necessarily complex.

Given a function $f\in L^{2}(G/K),$ let $\hat{f}$ denote the Helgason
Fourier transform of $f$, so that $\hat{f}$ is a square-integrable function
on $\mathfrak{a}^{\ast }\times B$ invariant under the action of the Weyl
group on $\mathfrak{a}^{\ast }.$ Here $B=K/M,$ where $M$ is the centralizer
of $\mathfrak{a}$ in $K.$ (See Section III.2 of \cite{He3}.) It is
convenient to think of $\hat{f}$ as a function on $\mathfrak{a}^{\ast }$
with values in $L^{2}(B).$ Thus for $\xi \in \mathfrak{a}^{\ast },$ we will
let $\left\Vert \hat{f}(\xi )\right\Vert $ be the $L^{2}$ norm of the
corresponding element of $L^{2}(B)$; that is, 
\begin{equation*}
\left\Vert \hat{f}(\xi )\right\Vert ^{2}=\int_{B}\left\vert \hat{f}(\xi
,b)\right\vert ^{2}db.
\end{equation*}%
The Plancherel theorem for the Fourier transform states that for $f\in
L^{2}(G/K)$%
\begin{equation}
\left\Vert f\right\Vert ^{2}=\int_{\mathfrak{a}^{\ast }}\left\Vert \hat{f}%
(\xi )\right\Vert ^{2}\frac{d\xi }{\left\vert c(\xi )\right\vert ^{2}}.
\label{plancherel}
\end{equation}%
Here $c$ is the Harish-Chandra $c$-function, the norm of $f$ is the $L^{2}$
norm with respect to the Riemannian volume measure on $G/K$, and $d\xi $
denotes the Lebesgue measure on $\mathfrak{a}^{\ast }$ (suitably normalized).

Meanwhile, let $\Delta $ denote the Laplacian on $G/K$, and let $e^{t\Delta
/2}$ denote the time-$t$ (forward) heat operator. (We take the Laplacian to
be a negative operator.) For $f\in L^{2}(G/K),$ let $F=e^{t\Delta /2}f.$ In
that case, $F$ is also in $L^{2}(G/K)$ and the Fourier transform of $F$ is
related to the Fourier transform of $f$ by%
\begin{equation}
\hat{F}(\xi )=e^{-t(\left\vert \xi \right\vert ^{2}+\left\vert \rho
\right\vert ^{2})/2}\hat{f}(\xi ),  \label{ff}
\end{equation}%
where $\rho $ is half the sum of the positive roots (with multiplicity).

According to Section 6 of \cite{KS2}, the function $F$ admits an analytic
continuation (also denoted $F$) to the domain $\Xi \subset G_{\mathbb{C}}/K_{%
\mathbb{C}}$ defined in (\ref{xi.def}). We now consider the integrals of $%
\left\vert F\right\vert ^{2}$ over various $G$-orbits inside $\Lambda .$ A
Gutzmer-type formula, due to J. Faraut \cite{Far1,Far2}, tells us that these
orbital integrals can computed as follows. Each $G$-orbit in $\Lambda $
contains exactly one point of the form $\exp _{x_{0}}(iZ),$ where $Z$
belongs to $\Omega ^{+}:=\Omega \cap \mathfrak{a}^{+}.$ Let $dg$ denote the
Haar measure on $G,$ normalized so that the push-forward of $dg$ to $G/K$
coincides with the Riemannian volume measure on $G/K.$ Then the Gutzmer
formula for $F$ takes the form (in light of (\ref{ff}))%
\begin{equation}
\int_{G}\left\vert F(g\cdot \exp _{x_{0}}(iY/2))\right\vert ^{2}dg=\int_{%
\mathfrak{a}^{\ast }}\left\Vert \hat{f}(\xi )\right\Vert
^{2}e^{-t(\left\vert \xi \right\vert ^{2}+\left\vert \rho \right\vert
^{2})}\phi _{\xi }(e^{iY})\frac{d\xi }{\left\vert c(\xi )\right\vert ^{2}},
\label{gutzmer}
\end{equation}%
for all $Y\in 2\Omega ^{+}.$ Here $\phi _{\xi }$ is the spherical function
normalized to equal 1 at $Y=0.$ Note that if $Y=0,$ then (\ref{gutzmer})
simply reduces to (\ref{plancherel}). Note also that on the left-hand side
of (\ref{gutzmer}) we have the $G$-orbit through the point $\exp
_{x_{0}}(iY/2),$ whereas on the right-hand side we have the spherical
function evaluated at $\exp (iY).$ This factor of 2 is the origin of the
factors of 2 in the isometry formula relative to the inversion formula. See
Appendix \ref{faraut.app} for more details about the Gutzmer formula and the
hypotheses under which it holds.

According to Lemma 2.1 of \cite{KOS}, for each $\xi \in \mathfrak{a}^{\ast
}, $ $\phi _{\xi }(iY)$ is defined and real-analytic for $Y\in 2\Omega .$
Furthermore, for a fixed $Y\in 2\Omega ,$ $\phi _{\xi }(e^{iY})$ grows at
most exponentially with $\xi ,$ with bounds that are uniform on each compact
subset of $2\Omega $. Thus, given $f\in L^{2}(G/K),$ the right-hand side of (%
\ref{gutzmer}) is a bounded as a function of $Y$ on each compact subset of $%
2\Omega .$

We now fix some bounded positive Ad-$K$-invariant density $\alpha $ on $%
\mathfrak{p}^{2R_{\max }}\subset T_{x_{0}}(G/K).$ Using the action of $G,$
we can identify every tangent space $T_{x}(G/K)$ with $\mathfrak{p},$ and
this identification is unique up to the adjoint action of $K$ on $\mathfrak{p%
}.$ Since $\alpha $ is Ad-$K$-invariant, we may unambiguously think of $%
\alpha $ as a function on each of the tangent spaces $T_{x}(G/K).$ We then
consider the integral%
\begin{equation}
G_{F}(R):=\int_{x\in G/K}\int_{Y\in T_{x}^{2R}(G/K)}\left\vert F(\exp
_{x}(iY/2))\right\vert ^{2}\alpha (Y)~dY~dx,  \label{rint}
\end{equation}%
where $T_{x}^{2R}(G/K)$ denotes the vectors in $T_{x}(G/K)$ with magnitude
less than $2R.$ As we shall see shortly, this integral will be well defined
and finite for all $R<R_{\max }.$

Now, for each $x\in G/K,$ we choose $g_{x}\in G$ so that $g_{x}\cdot
x_{0}=x, $ and we arrange for $g_{x}$ to be a measurable function of $x.$
(We may take, for example, $g_{x}\in P:=\exp \mathfrak{p}.$) Then we obtain
a measurable trivialization of the tangent bundle, with each tangent space $%
T_{x}(G/K)$ identified with $\mathfrak{p}=T_{x_{0}}(G/K)$ by means of the
action of $g_{x}.$ The integral in (\ref{rint}) then becomes an integral
over $(G/K)\times \mathfrak{p}^{2R},$ where $\mathfrak{p}^{2R}$ denotes the
set of points in $\mathfrak{p}$ with magnitude less than $2R.$ We now use
generalized polar coordinates to change the integration over $\mathfrak{p}%
^{2R}$ into one over $\mathfrak{a}_{2R}^{+}\times K$, where $\mathfrak{a}%
_{2R}^{+}=\mathfrak{a}^{+}\cap \mathfrak{p}^{2R}.$ This gives, after
applying Fubini's Theorem,%
\begin{equation}
G_{F}(R)=\int_{\mathfrak{a}_{2R}^{+}}\int_{G/K}\int_{K}\left\vert F(\exp
_{x}(i\mathrm{Ad}_{k}Y/2))\right\vert ^{2}~dk~dx~\alpha (Y)\mu (Y)~dY,
\label{rint2}
\end{equation}%
where $\mu $ is the density appearing in the generalized polar coordinates
(e.g., \cite[Thm. I.5.17]{He2}).

Since each coset $x$ in $G/K$ contains a unique element of the form $g_{x},$
each element $g$ of $G$ can be decomposed uniquely as $g=g_{x}k$, where $%
x=g\cdot x_{0}=gK$ and $k$ is an element of $K.$ In this way, we can
identify $G$ measurably with $(G/K)\times K.$ Let us consider the measure $%
dx~dk$ on $(G/K)\times K,$ where $dx$ denotes the Riemannian volume measure
and $dk$ is the normalized Haar measure on $K.$ If we transfer this measure
to $G$ by the above identification, the resulting measure on $G$ is
invariant under the left action of $G.$ To see this, note that for $h\in G$
and $x\in G/K,$ there exists a unique $k_{h,x}\in K$ such that $%
hg_{x}=g_{h\cdot x}k_{h,x}.$ Thus, the left action of $G$ on itself,
transferred to $(G/K)\times K,$ corresponds to the map $(x,k)\rightarrow
(h\cdot x,k_{h,x}k)$, and this action preserves $dx~dk.$ Thus, $dx~dk$
corresponds, under our identification, to a Haar measure $dg$ on the
(unimodular) group $G.$ Furthermore, by considering the case $Y=0$ in the
Gutzmer formula (\ref{gutzmer}), we can see that this Haar measure is
normalized the same way as the one in the Gutzmer formula.

Now, we have identified $T_{x}(G/K)$ with $\mathfrak{p}$ in such a way that $%
g_{x}\cdot (x_{0},Y)=(x,Y).$ Since the map $\Phi $ in (\ref{phi.map})
intertwines the action of $G$ on $\Lambda \subset T(G/K)$ with the action of 
$G$ on $\Xi \subset G_{\mathbb{C}}/K_{\mathbb{C}},$ we have that $g_{x}\cdot
\exp _{x_{0}}(iY/2)=\exp _{x}(iY/2)$ for all $Y\in \mathfrak{p}.$ Thus,%
\begin{equation*}
(g_{x}k)\cdot \exp _{x_{0}}(iY/2)=g_{x}\cdot \exp _{x_{0}}(i\mathrm{Ad}%
_{k}Y/2)=\exp _{x}(i\mathrm{Ad}_{k}Y/2).
\end{equation*}%
This means that the integrals over $G/K$ and over $K$ in (\ref{rint2})
combine into an integral over a $G$-orbit, giving%
\begin{equation}
G_{F}(R)=\int_{\mathfrak{a}_{2R}^{+}}\int_{G}\left\vert F(g\cdot \exp
_{x_{0}}(iY/2))\right\vert ^{2}dg~\alpha (Y)\mu (Y)~dY.  \label{gr.orbit}
\end{equation}%
\newline

We may then evaluate the integral over the $G$-orbits by Faraut's
Gutzmer-type formula (\ref{gutzmer}). After another application of Fubini's
Theorem, this gives%
\begin{equation}
G_{F}(R)=\int_{\mathfrak{a}}\left\Vert \hat{f}(\xi )\right\Vert
^{2}e^{-t(\left\vert \xi \right\vert ^{2}+\left\vert \rho \right\vert ^{2})}%
\left[ \int_{\mathfrak{a}_{2R}^{+}}\phi _{\xi }(e^{iY})\mu (Y)\alpha (Y)~dY%
\right] ~\frac{d\xi }{\left\vert c(\xi )\right\vert ^{2}}.  \label{gr.faraut}
\end{equation}%
We now use polar coordinates in the opposite direction to turn the integral
in square brackets back into an integral over $\mathfrak{p}^{2R}$:%
\begin{equation*}
\int_{\mathfrak{a}_{2R}^{+}}\phi _{\xi }(e^{iY})\mu (Y)\alpha (Y)~dY=\int_{%
\mathfrak{p}^{2R}}\phi _{\xi }(e^{iY})\alpha (Y)~dY.
\end{equation*}%
Since, as we have noted earlier, $\phi _{\xi }(iY)$ grow at most
exponentially as a function of $\xi $ with $Y$ fixed, with estimates that
are locally uniform in $Y$ (Lemma 2.1 of \cite{KOS}), it follows that $%
G_{F}(R)$ is finite for all $R<R_{\max }.$ (The growth of the quantity in
square brackets on the right-hand side of (\ref{gr.faraut}) is less rapid
than the decay of $\exp [-t(\left\vert \xi \right\vert ^{2}+\left\vert \rho
\right\vert ^{2})].$)

We have established, then, the following result.

\begin{proposition}
\label{partial.prop}For $f\in L^{2}(G/K)$ ($G$ not necessarily complex), let 
$F=e^{t\Delta /2}f$ and let $\alpha $ be a bounded, Ad-$K$-invariant,
positive density on $\mathfrak{p}^{2R_{\max }}.$ Then for all $R<R_{\max }$
the function%
\begin{equation*}
G_{F}(R):=\int_{x\in G/K}\int_{Y\in T_{x}^{2R}(G/K)}\left\vert F(\exp
_{x}(iY/2))\right\vert ^{2}\alpha (Y)~dY~dx
\end{equation*}%
is well-defined and finite and given by%
\begin{equation}
G_{F}(R)=\int_{\mathfrak{a}}\left\Vert \hat{f}(\xi )\right\Vert
^{2}e^{-t(\left\vert \xi \right\vert ^{2}+\left\vert \rho \right\vert ^{2})}%
\left[ \int_{\mathfrak{p}^{2R}}\phi _{\xi }(e^{iY})\alpha (Y)~dY\right] ~%
\frac{d\xi }{\left\vert c(\xi )\right\vert ^{2}}.  \label{partialprop.eq}
\end{equation}
\end{proposition}

Clearly, the quantity in square brackets on the right-hand side of (\ref%
{partialprop.eq}),%
\begin{equation}
\int_{\mathfrak{p}^{2R}}\phi _{\xi }(e^{iY})\alpha (Y)~dY,  \label{square}
\end{equation}%
is of vital importance in understanding Proposition \ref{partial.prop}. We
call this result a \textquotedblleft partial\textquotedblright\ isometry
formula, in that it involves integration of $\left\vert F(\exp
_{x}(iY)\right\vert ^{2}$ only over a tube of finite radius in $T(G/K).$ The
\textquotedblleft global\textquotedblright\ isometry formula, established in
Section \ref{globalisom.sec} in the complex case, will involve a (suitably
interpreted) limit of such partial isometries as the radius $R$ goes to
infinity.

To close this section, we wish to discuss why it is necessary to let the
radius tend to infinity. (Compare Section 4 of \cite{KOS}.) The goal, in the
end, is to have the right-hand side of (\ref{partialprop.eq}) be equal to $%
\left\Vert f\right\Vert ^{2}.$ To achieve greater flexibility in obtaining
this goal, we could replace $\mathfrak{p}^{2R}$ by any convex $K$-invariant
set in $\mathfrak{p}$ whose intersection with $\mathfrak{a}$ is contained in
the domain $2\Omega $. The largest such domain is $\Gamma :=\mathrm{Ad}%
_{K}(2\Omega ).$ Even if we replace $\mathfrak{p}^{2R}$ by $\Gamma ,$ the
evidence strongly suggests that there does exist any Ad-$K$-invariant
density $\alpha $ on $\Gamma $ for which the right-hand side of (\ref%
{partialprop.eq}) is equal to $\left\Vert f\right\Vert ^{2}.$

In order to have (\ref{partialprop.eq}) equal to $\left\Vert f\right\Vert
^{2}$ for all $f,$ $\alpha $ would have to satisfy%
\begin{equation}
\int_{\Gamma }\phi _{\xi }(e^{iY})\alpha (Y)~dY=e^{t(\left\vert \xi
\right\vert ^{2}+\left\vert \rho \right\vert ^{2})}  \label{impossible}
\end{equation}%
for almost every $\xi .$ (Essentially the same condition was obtained in a
slightly different way by Kr\"{o}tz, \'{O}lafsson, and Stanton in \cite[Eq.
4.29]{KOS}.) At least in the complex case (but almost certainly also in
general), a weight satisfying (\ref{impossible}) does not exist, as
demonstrated in Section 4 of \cite{KOS}.

Let us consider, for example, the case of hyperbolic 3-space. Then $\Gamma $
is just a ball of radius $\pi $ and the explicit formulas for the spherical
functions turns (\ref{impossible}) into%
\begin{equation}
\int_{\substack{ Y\in \mathbb{R}^{3}  \\ \left\vert Y\right\vert \leq \pi }}%
\frac{\sinh (\xi \left\vert Y\right\vert )}{\xi \sin \left\vert Y\right\vert 
}\alpha (Y)~dY=e^{t(\left\vert \xi \right\vert ^{2}+\left\vert \rho
\right\vert ^{2})},\quad \xi \in \mathbb{R}.  \label{impossible2}
\end{equation}%
Suppose $\alpha $ is any non-negative, rotationally invariant density for
which the left-hand side of (\ref{impossible2}) is finite for almost all $%
\xi .$ Then it is not hard to see that the left-hand side of (\ref%
{impossible2}) grows at most like $e^{\pi \left\vert \xi \right\vert },$ and
thus cannot equal the right-hand side of (\ref{impossible2}). A similar
argument applies to all symmetric spaces of the complex type, as explained
in \cite[Sect. 4]{KOS}.

This argument shows that (at least in the complex case), it is not possible
to express $\left\Vert f\right\Vert ^{2}$ as a $G$-invariant integral of $%
\left\vert F\right\vert ^{2}$ over the domain $\Xi .$ Thus, to obtain our
isometry formula in the complex case, we extend the integration beyond $\Xi $%
, using analytic continuation and a cancellation of singularities, as
explained in Section \ref{globalisom.sec}.

\section{ Strategy for a global isometry formula}

If we work by analogy to the results of Hall \cite{H1,H2} and Stenzel \cite%
{St1} in the compact case (see Theorem \ref{compact.thm} in the
introduction), then we want to take $\alpha $ to be something related to the
heat kernel for the \textit{compact} symmetric space $U/K$ dual to $G/K.$
Specifically, according to \cite{LGS,St1}, there is a natural local
identification of the fibers in $T(G/K)$ with the dual compact symmetric
space $U/K.$ We would like, if possible, to choose $\alpha $ so that $\alpha
(Y)dY$ is the heat kernel \textit{measure} on $U/K,$ based at the identity
coset and evaluated at time $2t$. More precisely, the results of \cite{HM3}
indicate that one should take $\alpha (Y)dY$ to be a sort of
\textquotedblleft unwrapped\textquotedblright\ version of this heat kernel
measure. (See Theorem 5 of \cite{HM3} and Section \ref{complexpartial.sec}
below for further discussion of the unwrapping concept.) This means that we
would like to take%
\begin{equation}
\alpha (Y)=\nu _{2t}^{\mathrm{c}}(Y)j^{\mathrm{c}}(Y),  \label{alpha.form}
\end{equation}%
where $\nu _{t}^{\mathrm{c}}$ is the unwrapped heat kernel \textit{density}
for $U/K$ and $j^{\mathrm{c}}$ is the Jacobian of the exponential mapping
for $U/K.$

With $\alpha $ as given above, the quantity in (\ref{square}) is given by%
\begin{equation}
\int_{\mathfrak{p}^{2R}}\phi _{\xi }(e^{iY})\alpha (Y)~dY=\int_{\mathfrak{p}%
^{2R}}\phi _{\xi }(e^{iY})\nu _{2t}^{\mathrm{c}}(Y)j^{\mathrm{c}}(Y)~dY.
\label{brackets}
\end{equation}%
Now, $\phi _{\xi }$ is an eigenfunction of the Laplacian on $G/K$ with
eigenvalue \thinspace $-(\left\vert \xi \right\vert ^{2}+\left\vert \rho
\right\vert ^{2}).$ It then follows that the the (locally defined) function
on $U/K$ given by $f(e^{Y})=\phi _{\xi }(e^{iY})$ is an eigenfunction of the
Laplacian for $U/K$ with eigenvalue $\left\vert \xi \right\vert
^{2}+\left\vert \rho \right\vert ^{2}.$ (This assertion can be verified by
direct computation but also follows from Theorem 1.16, Proposition 1.17,
Proposition 1.19 and Theorem 8.5 of \cite{LGS}.) If, by letting $R$ tend to
infinity, we could somehow make Proposition \ref{partial.prop} into a global
result (with $\alpha $ given by (\ref{alpha.form})), then we would be
integrating an eigenfunction of the Laplacian for $U/K$ against the heat
kernel for $U/K.$ Thus, the limit as $R$ tends to infinity of (\ref{brackets}%
) \textquotedblleft ought\textquotedblright\ to be $e^{t(\left\vert \xi
\right\vert ^{2}+\left\vert \rho \right\vert ^{2})}\phi _{\xi }(x_{0}).$
Since the spherical functions are normalized so that $\phi _{\xi }(x_{0})=1,$
we would get that the right-hand side of (\ref{partialprop.eq}) tends to $%
\left\Vert f\right\Vert ^{2}$ as $R$ tends to infinity.

If we could actually implement this program, we would then obtain an
isometry formula analogous to the one in the compact case: $\left\Vert
f\right\Vert ^{2}$ would be equal to the integral of $\left\vert
F\right\vert ^{2}$ first over the fibers with respect to the heat kernel
measure for the dual symmetric space and then over the base with respect to
the Riemannian volume measure. Unfortunately, because of the singularities
that occur in the analytically continued spherical functions and because the
identification of $\mathfrak{p}$ with $U/K$ is only local, we do not know
how to carry out the above strategy in general.

By contrast, J. Faraut has shown, using a Gutzmer-type formula due to
Lasalle \cite{Las}, that one can carry out a similar line of reasoning if
one \textit{starts} on a compact symmetric space. This leads \cite{Far3} to
a new proof of Stenzel's isometry formula for compact symmetric spaces.

In the noncompact case, the case in which $G$ is complex is the most
tractable one and we now specialize to this case. We will first work out
very explicitly the partial isometry formula in this case, by evaluating the
quantity in square brackets in (\ref{partialprop.eq}), with $\alpha $ given
by (\ref{alpha.form}). Then we let the radius tend to infinity, using an
appropriate cancellation of singularities.

\section{Partial isometry in the complex case\label{complexpartial.sec}}

We now assume that $G$ is a connected \textit{complex} semisimple group and $%
K$ a maximal compact subgroup. The assumption that $G$ is complex is
equivalent to the assumption that the (restricted) roots for $(G,K)$ form a
reduced root system with all roots having multiplicity 2. The complex case
is nothing but the noncompact dual to the compact group case studied in \cite%
{H1,H2}.

We make use of several (closely related) results that are specific to the
complex case and do not hold for general symmetric spaces of the noncompact
type. First, in the complex case, the dual compact symmetric space $U/K$ is
isometric to a compact group with a bi-invariant metric. There is, as a
result, a particular simple formula for the heat kernel on $U/K$, due to 
\`{E}skin \cite{E}. (See also \cite{U}.) We use an \textquotedblleft
unwrapped\textquotedblright\ version of the heat kernel density on $U/K,$
given by%
\begin{equation}
\nu _{2t}^{\mathrm{c}}(Y)=e^{t\left\vert \rho \right\vert ^{2}}j^{\mathrm{c}%
}(Y)^{-1/2}\frac{e^{-\left\vert Y\right\vert ^{2}/4t}}{(4\pi t)^{d/2}}.
\label{nu2t.form}
\end{equation}

This means that we want to take $\alpha $ in Proposition \ref{partial.prop}
to be (as in \ref{alpha.form})%
\begin{equation}
\alpha (Y)=\nu _{2t}^{\mathrm{c}}(Y)j^{\mathrm{c}}(Y)=e^{t\left\vert \rho
\right\vert ^{2}}j^{\mathrm{c}}(Y)^{1/2}\frac{e^{-\left\vert Y\right\vert
^{2}/4t}}{(4\pi t)^{d/2}},  \label{alpha.form2}
\end{equation}%
where on $\mathfrak{a}$ we have, explicitly, 
\begin{equation}
j^{\mathrm{c}}(Y)^{1/2}=\prod_{\alpha \in R^{+}}\frac{\sin \alpha (Y)}{%
\alpha (Y)}.  \label{jc}
\end{equation}%
As shown in \cite[Thm. 5]{HM3}, the signed measure $\nu _{2t}^{\mathrm{c}%
}(Y)j^{\mathrm{c}}(Y)~dY$ on $\mathfrak{p}$ is an \textquotedblleft
unwrapped\textquotedblright\ version of the heat kernel measure for $U/K.$
This means that the push-forward of this measure by $\exp :\mathfrak{p}%
\rightarrow U/K$ is precisely the heat kernel measure on $U/K$ at time $2t,$
based at the identity coset.

With $\alpha $ given by (\ref{alpha.form2}), the expression in (\ref{square}%
) is given by%
\begin{equation}
e^{t\left\vert \rho \right\vert ^{2}}\int_{\mathfrak{p}^{2R}}\phi _{\xi
}(e^{iY})j^{\mathrm{c}}(Y)^{1/2}\frac{e^{-\left\vert Y\right\vert ^{2}/4t}}{%
(4\pi t)^{d/2}}~dY.  \label{alphaint.complex}
\end{equation}%
Our next task is to compute (\ref{alphaint.complex}) as explicitly as
possible. Although there is an explicit formula for $\phi _{\xi }$ in the
complex case (see (\ref{spherical.form}) in Section \ref{kos.sec}), it is
not quite straightforward to compute (\ref{alphaint.complex}) using that
formula. We use instead a more geometric argument, which will also be useful
in studying the Segal--Bargmann transform on compact quotients of symmetric
spaces of the complex type.

It is known that the function $\phi _{\xi }$ is an eigenfunction for the
(non-Euclidean) Laplacian on $G/K$ with eigenvalue $-(\left\vert \xi
\right\vert ^{2}+\left\vert \rho \right\vert ^{2}).$ In the complex case, we
have special \textquotedblleft intertwining formulas\textquotedblright\ for
the Laplacian; see Proposition V.5.1 in \cite{He3} and the calculations for
the complex case on p. 484. These formulas tell us that the function $%
Y\rightarrow \phi _{\xi }(e^{Y})j^{\mathrm{nc}}(Y)^{1/2}$ is an
eigenfunction of the \textit{Euclidean} Laplacian for $\mathfrak{p}$ with
eigenvalue $-\left\vert \xi \right\vert ^{2}.$ (Here $j^{\mathrm{nc}}$ is
the Jacobian of the exponential mapping for the noncompact symmetric space $%
G/K.$) Since $j^{\mathrm{nc}}(iY)=j^{\mathrm{c}}(Y)$ (as is easily verified
from the formulas for these Jacobians) we see that the function%
\begin{equation}
\Psi _{\xi }(Y):=\phi _{\xi }(e^{iY})j^{\mathrm{c}}(Y)^{1/2}  \label{psi.xi}
\end{equation}%
is an eigenfunction of the Euclidean Laplacian on $\mathfrak{p}^{R}$ with
eigenvalue $\left\vert \xi \right\vert ^{2}.$

\begin{lemma}
\label{euclideanint.lem}Let $\Psi $ be a smooth function on the ball $%
B(2R_{0},0)$ in $\mathbb{R}^{d}$ satisfying $\Delta \Psi =\sigma \Psi $ for
some constant $\sigma \in \mathbb{R},$ where $\Delta $ is the Euclidean
Laplacian. Let $\beta $ be a non-negative, bounded, measurable, rotationally
invariant function on $B(2R_{0},0).$ Then for all $R<R_{0}$ we have%
\begin{equation}
\int_{\left\vert Y\right\vert \leq 2R}\Psi (Y)\beta (Y)~dY=\Psi
(0)\int_{\left\vert Y\right\vert \leq 2R}e^{\sqrt{\sigma }y_{1}}\beta (Y)~dY.
\label{euclidean.eq}
\end{equation}%
Here $Y=(y_{1},\ldots ,y_{d})$ and $\sqrt{\sigma }$ is either of the two
square roots of $\sigma .$
\end{lemma}

\begin{proof}
We let $\tilde{\Psi}$ denote the radialization of $\Psi $ in the Euclidean
sense, that is, the average of $\Psi $ with respect to the action of the
rotation group. Then $\tilde{\Psi}$ is also an eigenfunction of the
Euclidean Laplacian with the same eigenvalue $\sigma ,$ and $\tilde{\Psi}%
(0)=\Psi (0).$ Since $\beta $ is rotationally invariant, replacing $\Psi $
with $\tilde{\Psi}$ does not change the value of the integral. But since $%
\tilde{\Psi}$ is radial, it satisfies differential equation%
\begin{equation}
\frac{d^{2}\tilde{\Psi}}{dr^{2}}+\frac{(d-1)}{r}\frac{d\tilde{\Psi}}{dr}%
=\sigma \tilde{\Psi},  \label{ode}
\end{equation}%
with $\tilde{\Psi}(0)$ finite and $\left. d\tilde{\Psi}/dr\right\vert
_{r=0}=0.$

When $d=1,$ the equation (\ref{ode}) is nonsingular at the origin and
standard uniqueness results show that $\tilde{\Psi}$ is determined by $%
\tilde{\Psi}(0)$. When $d\geq 2,$ (\ref{ode}) is a second-order, linear,
nonconstant-coefficient equation, with a regular singular point at $r=0$. A
simple calculation with the theory of regular singular points shows that
there is, up to a constant, only one solution of this equation that is
nonsingular at the origin.

Now let $\gamma (Y)=e^{\sqrt{\sigma }y_{1}},$ which is also an eigenfunction
of the Laplacian with eigenvalue $\sigma .$ If $\tilde{\gamma}$ denotes the
Euclidean radialization of $\gamma ,$ then $\tilde{\gamma}(0)=1$ and $\tilde{%
\gamma}$ also solves the equation (\ref{ode}) above. Thus we must have $%
\tilde{\Psi}=\tilde{\Psi}(0)\tilde{\gamma}=\Psi (0)\tilde{\gamma}.$ So in
the integral on the left-hand side of (\ref{euclidean.eq}) we may replace $%
\Psi $ by $\tilde{\Psi}$ and then by $\tilde{\Psi}(0)\tilde{\gamma}$ and
finally by $\Psi (0)\gamma ,$ which establishes the lemma.
\end{proof}

We are now ready to put everything together. We apply Proposition \ref%
{partial.prop} with $\alpha $ as given in (\ref{alpha.form2}). We make use
of Lemma \ref{euclideanint.lem} with $\beta (Y)$ equal to $(4\pi
t)^{-d/2}\exp (-\left\vert Y\right\vert ^{2}/4t)$, $\Psi $ equal to the
function $\Psi _{\xi }$ in (\ref{psi.xi}), and $\sigma $ equal to $%
\left\vert \xi \right\vert ^{2}.$ We also make the change of variable $%
Y\rightarrow 2Y$ (for cosmetic reasons) in the integral that defines $%
G_{F}(R).$ The result is the following.

\begin{theorem}[Partial Isometry Formula]
\label{partial.thm}Let $f$ be in $L^{2}(G/K)$ ($G$ complex) and let $%
F=e^{t\Delta /2}f$. Then for all $R<R_{\max }$ the function $G_{F}(R)$
defined by%
\begin{equation*}
G_{F}(R)=\int_{x\in G/K}\int_{Y\in T_{x}^{R}(G/K)}\left\vert F(\exp
_{x}(iY))\right\vert ^{2}\nu _{2t}^{\mathrm{c}}(2Y)j^{\mathrm{c}%
}(2Y)~2^{d}dY~dx
\end{equation*}%
may be computed as%
\begin{equation}
G_{F}(R)=\int_{\mathfrak{a}}\left\Vert \hat{f}(\xi )\right\Vert
^{2}e^{-t(\left\vert \xi \right\vert ^{2}+\left\vert \rho \right\vert ^{2})}%
\left[ e^{t\left\vert \rho \right\vert ^{2}}\int_{\substack{ Y\in \mathbb{R}%
^{d}  \\ \left\vert Y\right\vert \leq 2R}}e^{\left\vert \xi \right\vert
y_{1}}\frac{e^{-\left\vert Y\right\vert ^{2}/4t}}{(4\pi t)^{d/2}}~dy\right] ~%
\frac{d\xi }{\left\vert c(\xi )\right\vert ^{2}},  \label{partial.eq}
\end{equation}%
where $Y=(y_{1},\ldots ,y_{d}).$ Here $T_{x}^{R}(G/K)$ is the set of vectors
in $T_{x}(G/K)$ with magnitude less than $R$ and $\nu _{2t}^{\mathrm{c}}$
and $j^{\mathrm{c}}$ are as in (\ref{nu2t.form}) and (\ref{jc}).
\end{theorem}

Note that for a given $R,$ the expression in square brackets on the
right-hand side of (\ref{partial.eq}) depends only on $\left\vert \xi
\right\vert $. Since the effect of the Laplacian on the Fourier transform of 
$f$ is to multiply $\hat{f}(\xi )$ by $-(\left\vert \xi \right\vert
^{2}+\left\vert \rho \right\vert ^{2}),$ we can rewrite (\ref{partial.eq})
as 
\begin{equation}
G_{F}(R)=\left\langle f,\beta _{R}(-\Delta )f\right\rangle _{L^{2}(G/K)},
\label{gr.funct}
\end{equation}
where $\beta _{R}$ is the function given by%
\begin{equation}
\beta _{R}(\lambda )=e^{-t\lambda }e^{t\left\vert \rho \right\vert ^{2}}\int 
_{\substack{ y\in \mathbb{R}^{d}  \\ \left\vert y\right\vert \leq 2R}}\exp
\left( \sqrt{\lambda -\left\vert \rho \right\vert ^{2}}~y_{1}\right) \frac{%
e^{-\left\vert y\right\vert ^{2}/4t}}{(4\pi t)^{d/2}}~dy  \label{beta.def}
\end{equation}%
Note that the $L^{2}$ spectrum of $-\Delta $ is $[\left\vert \rho
\right\vert ^{2},\infty )$, so that the argument of the square root on the
right-hand side of (\ref{beta.def}) is always non-negative.

\section{Global isometry in the complex case\label{globalisom.sec}}

Our goal now is to \textquotedblleft let $R$ tend to
infinity\textquotedblright\ in our partial isometry formula for the complex
case (Theorem \ref{partial.thm}). That it is possible to do so reflects a 
\textit{cancellation of singularities}. The function $F(\exp _{x}iY)$
becomes singular (and multiple-valued) for large $Y.$ Reflecting this, the
orbital integrals of $\left\vert F\right\vert ^{2}$ become unbounded as the
orbits approach the boundary of the domain $\Xi .$ However, Faraut's
Gutzmer-type formula tells us that the singularities in the orbital
integrals are controlled by the singularities in the analytically continued
spherical functions. In the complex case, the singularities of the
analytically continued spherical functions are of a particularly simple sort
(see (\ref{spherical.form}) in Section \ref{kos.sec}). These singularities
are cancelled by the zeros in the density against which we are integrating
the orbital integrals, namely, the function $\alpha $ given in (\ref%
{alpha.form2}). (Compare (\ref{jc}) to (\ref{spherical.form}).) This
cancellation of singularities allows $G_{F}(R)$ to be nonsingular, even
though both $F$ itself and the orbital integrals of $\left\vert F\right\vert
^{2}$ are singular.

In the proof of Theorem \ref{partial.thm}, the above-described cancellation
of singularities is reflected in the fact that the expression in square
brackets on the right-hand side of (\ref{partial.eq}) is well defined and
finite for all $R.$ It is not hard, then, to show that $G_{F}(R)$ admits a
real-analytic extension to the whole positive half-line. Furthermore, the
limit as $R$ tends to infinity of this analytic extension is easily
evaluated by setting $R=\infty $ on the right-hand side of (\ref{partial.eq}%
) and evaluating a standard Gaussian integral. This will lead to the
following result.

\begin{theorem}[Global Isometry Formula]
\label{globalisom.thm}Let $f$ be in $L^{2}(G/K),$ with $G$ complex, and let $%
F=e^{t\Delta /2}f.$ Then for all $R<R_{\max },$ the quantity 
\begin{equation*}
G_{F}(R):=\int_{x\in G/K}\int_{Y\in T_{x}^{R}(G/K)}\left\vert F(\exp
_{x}(iY))\right\vert ^{2}\nu _{2t}^{\mathrm{c}}(2Y)j^{\mathrm{c}%
}(2Y)~2^{d}dY~dx
\end{equation*}%
is defined and finite. Furthermore, the function $G_{F}$ has a real-analytic
extension from $(0,R_{\max })$ to $(0,\infty )$ and this extension (also
denoted $G_{F}$) satisfies 
\begin{equation*}
\lim_{R\rightarrow \infty }G_{F}(R)=\left\Vert f\right\Vert
_{L^{2}(G/K)}^{2}.
\end{equation*}
\end{theorem}

\begin{proof}
We consider the right-hand side of (\ref{partial.eq}) and wish to show that
this expression is finite for all $R\in (0,\infty )$ and that it is
real-analytic in $R.$ The quantity in square brackets in (\ref{partial.eq})
is bounded by its limit as $R$ tends to infinity, which is equal to $%
e^{t(\left\vert \xi \right\vert ^{2}+\left\vert \rho \right\vert ^{2})}.$
(This is a simple Gaussian integral.) Thus the right-hand side of (\ref%
{partial.eq}) is bounded by%
\begin{equation*}
\int_{\mathfrak{a}}\left\Vert \hat{f}(\xi )\right\Vert ^{2}\frac{d\xi }{%
\left\vert c(\xi )\right\vert ^{2}}=\left\Vert f\right\Vert ^{2}<\infty .
\end{equation*}

To see that the right-hand side of (\ref{partial.eq}) is real-analytic as a
function of $R,$ we reverse the order of integration (since everything is
positive) and write it as%
\begin{equation}
\int_{\substack{ y\in \mathbb{R}^{d}  \\ \left\vert y\right\vert \leq 2R}}%
\left[ \int_{\mathfrak{a}}\left\Vert \hat{f}(\xi )\right\Vert
^{2}e^{-t(\left\vert \xi \right\vert ^{2}+\left\vert \rho \right\vert
^{2})}e^{t\left\vert \rho \right\vert ^{2}}e^{\left\vert \xi \right\vert
y_{1}}\frac{d\xi }{\left\vert c(\xi )\right\vert ^{2}}\right] \frac{%
e^{-\left\vert y\right\vert ^{2}/4t}}{(4\pi t)^{d/2}}~dy.
\label{rhs.reverse}
\end{equation}%
Now, for any complex number $y_{1},$ the quantity $e^{-t\left\vert \xi
\right\vert ^{2}}e^{\left\vert \xi \right\vert y_{1}}$ is a bounded function
of $\xi $. It is therefore not hard to see (using Morera's Theorem) that the
expression in square brackets in (\ref{rhs.reverse}) admits an extension
(given by the same formula) to an entire function of $y_{1}.$ It follows
that the whole integrand in (\ref{rhs.reverse}) is a real-analytic function
of $y.$ It is then a straightforward exercise to verify that the integral of
a real-analytic function over a ball of radius $R$ is a real-analytic
function of $R.$

To evaluate the limit as $R$ tends to infinity of the right-hand side of (%
\ref{partial.eq}), we use monotone convergence to put the limit inside. The
quantity in square brackets then becomes an easily evaluated Gaussian
integral:%
\begin{equation}
e^{t\left\vert \rho \right\vert ^{2}}\int_{y\in \mathbb{R}^{d}}e^{\left\vert
\xi \right\vert y_{1}}\frac{e^{-\left\vert y\right\vert ^{2}/4t}}{(4\pi
t)^{d/2}}~dy=e^{t\left\vert \rho \right\vert ^{2}}e^{t\left\vert \xi
\right\vert ^{2}}.  \label{gauss_int}
\end{equation}%
Thus, the right-hand side of (\ref{partial.eq}) converges as $R$ tends to
infinity to 
\begin{equation*}
\int_{\mathfrak{a}}\left\Vert \hat{f}(\xi )\right\Vert ^{2}d\xi /\left\vert
c(\xi )\right\vert ^{2}=\left\Vert f\right\Vert _{L^{2}(G/K)}^{2},
\end{equation*}%
which is what we want.
\end{proof}

\section{Surjectivity theorem in the complex case\label{surjectivity.sec}}

Our goal is to show that if $F$ is any holomorphic function for which the
isometry formula makes sense and is finite, then $F$ is the analytic
continuation of $e^{t\Delta /2}f,$ for some unique $f\in L^{2}(G/K).$ In
contrast to the surjectivity result in \cite{KOS}, we do not assume that the
restriction of $F$ to $G/K$ is in $L^{2}(G/K)$ with rapidly decaying Fourier
transform. Rather, this property of $F$ holds automatically, in light of the
strong form of the Gutzmer formula established in \cite{Far2}. (See also
Appendix \ref{faraut.app}.)

\begin{theorem}
\label{surjectivity.thm}Suppose $F$ is a holomorphic function on a domain of
the form%
\begin{equation}
\left\{ \left. \exp _{x}(iY)\in \Xi \right\vert (x,Y)\in
T^{R_{0}}(G/K)\right\}  \label{r0domain}
\end{equation}%
for some $R_{0}\leq R_{\max }$ and suppose that the function%
\begin{equation}
G_{F}(R):=\int_{x\in G/K}\int_{Y\in T_{x}^{2R}(G/K)}\left\vert F(\exp
_{x}(iY))\right\vert ^{2}\nu _{2t}^{\mathrm{c}}(2Y)j^{\mathrm{c}%
}(2Y)~2^{d}dY~dx  \label{grdef}
\end{equation}%
is finite for all sufficiently small $R.$ Suppose further that $G_{F}$ has a
real-analytic extension to $(0,\infty )$ and that%
\begin{equation*}
\lim_{R\rightarrow \infty }G_{F}(R)
\end{equation*}%
exists and is finite. Then there exists a unique $f\in L^{2}(G/K)$ with $%
\left. F\right\vert _{G/K}=e^{t\Delta /2}f.$
\end{theorem}

Although we initially assume that $F$ is holomorphic only on a domain of the
form (\ref{r0domain}), after the fact we see that the function $F$, being
the analytic continuation of a function of the form $e^{t\Delta /2}f,$ can
be extended holomorphically to all of $\Xi .$ Furthermore, once $%
F=e^{t\Delta /2}f,$ the isometry theorem tells us that the limit as $%
R\rightarrow \infty $ of $G_{F}(R)$ is $\left\Vert f\right\Vert ^{2}.$

\begin{proof}
The uniqueness of $f$ follows from the injectivity of the heat operator $%
e^{t\Delta /2},$ which in turn follows from the spectral theorem or from the
Fourier transform or from the isometry formula.

We turn now to proving the existence of $f.$ According to results of Faraut 
\cite{Far2}, the assumption that $F$ is square-integrable over the domain in
(\ref{r0domain}) implies that the restriction of $F$ to $G/K$ is in $%
L^{2}(G/K),$ that the orbital integrals of $\left\vert F\right\vert ^{2}$
inside this domain are finite, and that these orbital integrals are given by
the Gutzmer formula (\ref{gutzmer}). Thus, if we compute the right-hand side
of (\ref{grdef}) by method of the previous section (as in the proof of (\ref%
{rhs.reverse})), we conclude that%
\begin{equation}
G_{F}(R)=\int_{\substack{ y\in \mathbb{R}^{d}  \\ \left\vert y\right\vert
\leq 2R}}\left[ \int_{\mathfrak{a}}\left\Vert \widehat{\left. F\right\vert
_{G/K}}\right\Vert ^{2}e^{t\left\vert \rho \right\vert ^{2}}e^{\left\vert
\xi \right\vert y_{1}}\frac{d\xi }{\left\vert c(\xi )\right\vert ^{2}}\right]
\frac{e^{-\left\vert y\right\vert ^{2}/4t}}{(4\pi t)^{d/2}}~dy  \label{gr1}
\end{equation}%
for $R<R_{0}.$

We now wish to show that the analytic continuation of $G_{F}$ must be given
(for all $R\in (0,\infty )$) by the expression on the right-hand side of (%
\ref{gr1}). Since the Gaussian factor on the right-hand side of (\ref{gr1})
is rotationally invariant, the whole integral is unchanged if we replace the
quantity in square brackets (viewed a function of $y$) by its average over
the action of the rotation group. This averaging can be put inside the
integral over $\mathfrak{a}$, at which point it affects only $e^{\left\vert
\xi \right\vert y_{1}}.$ Averaging this function gives (after interchanging
an integral with a uniformly convergent sum)%
\begin{equation}
\sum_{n~\mathrm{even}}^{{}}\frac{1}{n!}\left\vert \xi \right\vert
^{n}\left\vert y\right\vert ^{n}\int_{S^{d}}(u\cdot e_{1})^{n}~du,
\label{exp.ave}
\end{equation}%
where $du$ is the normalized volume measure on $S^{d}$ and where the terms
for $n$ odd are zero.

If we replace $e^{\left\vert \xi \right\vert y_{1}}$ by (\ref{exp.ave}) on
the right-hand side of (\ref{gr1}), all quantities involved will be
positive, so by Fubini's Theorem we may freely rearrange the sums and
integrals. Rearranging and using polar coordinates on the integral over $%
\mathbb{R}^{d}$ gives 
\begin{equation}
G_{F}(R)=\int_{0}^{2R}\left[ \sum_{n~\mathrm{even}}^{{}}r^{n}\int_{S^{d}}(u%
\cdot e_{1})^{n}~du\int_{\mathfrak{a}}\left\Vert \widehat{\left.
F\right\vert _{G/K}}\right\Vert ^{2}e^{t\left\vert \rho \right\vert
^{2}}\left\vert \xi \right\vert ^{n}\frac{d\xi }{\left\vert c(\xi
)\right\vert ^{2}}\right] \frac{e^{-r^{2}/4t}}{(4\pi t)^{d/2}}%
c_{d}r^{d-1}~dr,  \label{gr2}
\end{equation}%
where $c_{d}$ is the volume of the unit sphere in $\mathbb{R}^{d}.$
Differentiating with respect to $R$ and moving some factors to the other
side gives%
\begin{equation}
G_{F}^{\prime }(R)(4\pi t)^{d/2}e^{R^{2}/4t}c_{d}^{-1}(2R)^{1-d}=2\sum_{n~%
\mathrm{even}}^{{}}(2R)^{n}\int_{S^{d}}(u\cdot e_{1})^{n}~du\int_{\mathfrak{a%
}}\left\Vert \widehat{\left. F\right\vert _{G/K}}\right\Vert
^{2}e^{t\left\vert \rho \right\vert ^{2}}\left\vert \xi \right\vert ^{n}%
\frac{d\xi }{\left\vert c(\xi )\right\vert ^{2}}  \label{gr.prime}
\end{equation}%
for $R<R_{0}.$

Now, since $G_{F}(R)$ admits a real-analytic extension to all of $(0,\infty
) $, so does the right-hand side of (\ref{gr.prime}). Since the coefficient
of $R^{n}$ in (\ref{gr.prime}) is non-negative for all $n,$ it follows (see
Lemma \ref{convergence} below) that the series on the right-hand side of (%
\ref{gr.prime}) must have infinite radius of convergence. Then both sides of
(\ref{gr.prime}) are defined and real-analytic for all positive $R;$ since
they are equal for small $R,$ they must be equal for all $R.$ It then
follows that (\ref{gr2}) also holds for all $R.$ Undoing the reasoning that
led to (\ref{gr2}), we conclude that (\ref{gr1}) also holds for all $R.$

Now that we know that the analytic continuation of $G_{F}(R)$ is given by (%
\ref{gr1}) for all $R,$ the Monotone Convergence Theorem tells us that%
\begin{equation*}
\lim_{R\rightarrow \infty }G_{F}(R)=\int_{\mathbb{R}^{d}}\left[ \int_{%
\mathfrak{a}}\left\Vert \widehat{\left. F\right\vert _{G/K}}(\xi
)\right\Vert ^{2}e^{t\left\vert \rho \right\vert ^{2}}e^{\left\vert \xi
\right\vert y_{1}}\frac{d\xi }{\left\vert c(\xi )\right\vert ^{2}}\right] 
\frac{e^{-\left\vert y\right\vert ^{2}/4t}}{(4\pi t)^{d/2}}~dy.
\end{equation*}%
Reversing the order of integration and using again the Gaussian integral (%
\ref{gauss_int}) will then give%
\begin{equation}
\int_{\mathfrak{a}}\left\Vert \widehat{\left. F\right\vert _{G/K}}(\xi
)\right\Vert ^{2}e^{t\left\vert \rho \right\vert ^{2}}e^{t\left\vert \xi
\right\vert ^{2}}\frac{d\xi }{\left\vert c(\xi )\right\vert ^{2}}%
=\lim_{R\rightarrow \infty }G(R)<\infty .  \label{ft}
\end{equation}%
We may then conclude that $\left. F\right\vert _{G/K}$ is of the form $%
e^{t\Delta /2}f,$ where $f$ is the function whose Fourier transform is given
by%
\begin{equation*}
\hat{f}(\xi )=\widehat{\left. F\right\vert _{G/K}}(\xi )e^{t(\left\vert \rho
\right\vert ^{2}+\left\vert \xi \right\vert ^{2})/2}.
\end{equation*}%
(That there really is an $L^{2}$ function $f$ with this Fourier transform
follows from (\ref{ft})).

This concludes the proof of surjectivity, except for the following
elementary lemma about power series with non-negative terms.
\end{proof}

\begin{lemma}
\label{convergence}Suppose $H$ is a real-analytic function on $(0,\infty )$
such that on $(0,\varepsilon ),$ $H$ is given by a convergent power series $%
H(R)=\sum_{n=0}^{\infty }a_{n}R^{n}.$ Suppose also the coefficients $a_{n}$
are non-negative. Then the series $\sum_{n=0}^{\infty }a_{n}R^{n}$ has
infinite radius of convergence.
\end{lemma}

\begin{proof}
Assume, to the contrary, that the series $\sum a_{n}x^{n}$ has radius of
convergence $S<\infty .$ Since both $H(R)$ and $\sum a_{n}R^{n}$ are real
analytic on $(0,S)$ and they are equal on $(0,\varepsilon ),$ they are equal
on $(0,S).$We may then differentiate $H(R)$ term by term for $R<S$. Since $%
H^{(k)}$ is continuous on $(0,\infty )$, letting $R$ approach $S$ gives, by
Monotone Convergence,%
\begin{equation*}
\frac{H^{(k)}(S)}{k!}=\sum_{n=0}^{\infty }a_{n}\binom{n}{k}S^{n-k},
\end{equation*}%
where $\binom{n}{k}$ is defined to be 0 for $k>n.$

Using Fubini's theorem (since all terms are non-negative) and the binomial
theorem, we have for any $\delta >0,$%
\begin{eqnarray*}
\sum_{k=0}^{\infty }\frac{H^{(k)}(S)}{k!}\delta ^{k} &=&\sum_{n=0}^{\infty
}a_{n}\sum_{k=0}^{\infty }\binom{n}{k}S^{n-k}\delta ^{k} \\
&=&\sum_{n=0}^{\infty }a_{n}(S+\delta )^{n}=\infty ,
\end{eqnarray*}%
because $\sum a_{n}R^{n}$ has radius of convergence $S.$ This shows that the
Taylor series of $H$ at $S$ has radius of convergence zero, contradicting
the assumption that $H$ is real-analytic on $(0,\infty ).$
\end{proof}

\section{Comparison with the results of Kr\"{o}tz, \'{O}lafsson, and Stanton 
\label{kos.sec}}

As we have already pointed out in Section \ref{kos.sub}, the isometry
formula of Kr\"{o}tz, \'{O}lafsson, and Stanton (Theorem 3.3 of \cite{KOS}),
when specialized to the complex case, \textit{does not} reduce to our
isometry formula. We now explain the relationship between the two formulas.
Since both formulas already have complete proofs, we will not attempt to
give a completely rigorous reduction of one formula to the other. Rather, we
will show formally how the isometry formula in \cite{KOS} can be reduced to
the one we prove here, by means of an integration by parts.

Let us begin in the setting of \cite{KOS}, which means that we consider a
symmetric space of the form $G/K$, where $G$ is a real connected semisimple
group with finite center and $K$ is a maximal compact subgroup. At the
moment, we do not assume that $G$ is complex. After adjusting for
differences of normalization of the heat operator ($e^{t\Delta }$ in \cite%
{KOS} versus $e^{t\Delta /2}$ here), the isometry formula of \cite[Thm. 3.3]%
{KOS} can be written as%
\begin{equation}
\left\Vert f\right\Vert ^{2}=\frac{e^{t|\rho |^{2}}}{\left\vert W\right\vert
(4\pi t)^{\frac{n}{2}}}\int_{\mathfrak{a}}D\left( \mathcal{O}_{\left\vert
F\right\vert ^{2}}(iY)\right) e^{-\left\vert Y\right\vert ^{2}/4t}dY,
\label{kos.eq}
\end{equation}%
where $n=\dim \mathfrak{a}$ is the rank of $G/K.$ Here $\mathcal{O}%
_{\left\vert F\right\vert ^{2}}(iY)$ denotes the \textquotedblleft orbital
integral\textquotedblright\ of $\left\vert F\right\vert ^{2}$ appearing in (%
\ref{gutzmer}), namely,%
\begin{equation}
\mathcal{O}_{\left\vert F\right\vert ^{2}}(iY)=\int_{G}\left\vert F(g\cdot
\exp _{x_{0}}(iY/2))\right\vert ^{2}dg  \label{orbital}
\end{equation}%
and $D$ is a pseudodifferential \textquotedblleft shift
operator\textquotedblright\ that takes the spherical functions to their
Euclidean counterparts. Although $\mathcal{O}_{\left\vert F\right\vert
^{2}}(iY)$ itself is defined only for small $Y,$ the shift operator $D$
cancels out all the singularities and produces a function that is defined
real-analytically on all of $\mathfrak{a}.$

(There appears to be a slight inconsistency in the way the orbital integral
is defined in \cite{KOS}, as in (\ref{orbital}) in the original definition,
but with $Y/2$ replaced by $Y$ in Equation (3.19) in the proof of the
isometry formula. We have maintained the original definition (Equation (1.2)
of \cite{KOS}) of the orbital integral and adjusted the isometry formula
accordingly. This adjustment along with the difference in normalization of
the heat equation account for the differences between (\ref{kos.eq}) and
Theorem 3.3 of \cite{KOS}.)

If we ignored the singularities in $\mathcal{O}_{\left\vert F\right\vert
^{2}},$ we could formally move the operator $D$ off of the orbital integral,
at the expense of applying the adjoint operator $D^{\ast }$ to the Gaussian
factor. We could then use Weyl invariance to reduce the domain of
integration from $\mathfrak{a}$ to $\mathfrak{a}^{+},$ giving the
nonrigorous expression%
\begin{equation}
\left\Vert f\right\Vert ^{2}\overset{?}{=}\frac{e^{t|\rho |^{2}}}{(4\pi t)^{%
\frac{n}{2}}}\int_{\mathfrak{a}^{+}}\mathcal{O}_{\left\vert F\right\vert
^{2}}(iY)D^{\ast }\!\left( e^{-\left\vert Y\right\vert ^{2}/4t}\right) dY.
\label{fake.norm}
\end{equation}

The idea is that $D^{\ast }$ is also a sort of shift operator (or Abel
transform) and should have the effect of turning the Euclidean heat kernel $%
\exp (-\left\vert Y\right\vert ^{2}/4t)$ into the non-Euclidean heat kernel
for the \textit{compact} symmetric space dual to $G/K.$ If (\ref{fake.norm})
were really correct it would express $\left\Vert f\right\Vert ^{2}$ as an
integral of $\left\vert F\right\vert ^{2}$ as an integral over $G$-orbits
followed by an integral over the space of $G$-orbits, which is just the sort
of thing we have in this paper.

In general, it is not at all clear that the right-hand side of (\ref%
{fake.norm}) makes sense. Even assuming that $D^{\ast }(\exp (-\left\vert
Y\right\vert ^{2}/4t))$ is well defined, there will be singularities in the
orbital integral $\mathcal{O}_{\left\vert F\right\vert ^{2}}(iY),$ which are
related to the singularities in the analytically continued spherical
functions that appear in the Gutzmer formula. Examples show that in general,
the singularities in the orbital integral will \textit{not} be canceled by
zeros in $D^{\ast }(\exp (-\left\vert Y\right\vert ^{2}/4t))$ and so the
right-hand side of (\ref{fake.norm}) will not be well defined without some
further \textquotedblleft interpretation.\textquotedblright

In the complex case, however, $D$ is a simple differential operator and
taking its adjoint amounts to integrating by parts. We will now compute $%
D^{\ast }$ explicitly and see that, in this case, $D^{\ast }(\exp
(-\left\vert Y\right\vert ^{2}/4t))$ has zeros in all the places that the
orbital integral is singular, so that (\ref{fake.norm}) is actually
nonsingular. Indeed, in the complex case, (\ref{fake.norm}) is essentially
just our isometry formula (Theorem \ref{globalisom.thm}).

In this calculation, there are various constants, depending only on the
choice of symmetric space, whose values are not worth keeping track of. In
the remainder of this section, $C$ will denote such a constant whose value
changes from line to line.

In the complex case, the explicit formula for the spherical function (e.g.,
Theorem 5.7, p. 432, of \cite{He3}) implies that%
\begin{equation}
\phi _{\xi }(e^{iY})=\frac{C}{\pi (\xi )}\cdot \frac{\sum_{w\in W}(\det
w)e^{-\left\langle w\cdot \xi ,Y\right\rangle }}{\Pi _{\alpha \in \Sigma
^{+}}\sin \left\langle \alpha ,Y\right\rangle },  \label{spherical.form}
\end{equation}%
where $\pi $ is the Weyl-alternating polynomial given by $\pi
(Y)=\prod_{\alpha \in \Sigma ^{+}}\left\langle \alpha ,Y\right\rangle .$
Meanwhile, $D$ is supposed to be the operator that takes the spherical
functions to their Euclidean counterparts $\psi _{\xi },$ which satisfy%
\begin{equation*}
\psi _{\xi }(iY)=\sum_{w\in W}e^{-\left\langle w\cdot \xi ,Y\right\rangle }.
\end{equation*}%
(Note that, following \cite{KOS}, we normalize the Euclidean spherical
functions to have the value $\left\vert W\right\vert $ at the origin.)

Let $D_{\alpha }$ denote the directional derivative in the direction of $%
\alpha $ and observe that 
\begin{equation*}
\left( \prod_{\alpha \in \Sigma ^{+}}(-D_{\alpha })\right) e^{-\left\langle
\xi ,w\cdot Y\right\rangle }=\left( \prod_{\alpha \in \Sigma
^{+}}\left\langle w\cdot \xi ,\alpha \right\rangle \right) e^{-\left\langle
\xi ,w\cdot Y\right\rangle }=(\det w)\pi (\xi )e^{-\left\langle \xi ,w\cdot
Y\right\rangle }
\end{equation*}%
because the polynomial $\pi $ is alternating. Thus, we can see that%
\begin{equation}
D=C\left( \prod_{\alpha \in \Sigma ^{+}}(-D_{\alpha })\right) \left(
\prod_{\alpha \in \Sigma ^{+}}\sin \left\langle \alpha ,Y\right\rangle
\right) .  \label{d.form}
\end{equation}%
(To be precise, the operator that we are here calling $D$ is the operator
that takes the \textit{analytic continuation} of the spherical function $%
\phi _{\xi }$ for $G/K$ to the \textit{analytic continuation} of the
Euclidean spherical function $\psi _{\xi }.$ The operator that takes $\phi
_{\xi }$ itself to $\psi _{\xi }$ would involve hyperbolic sines instead of
ordinary sines. With our definition of $D,$ it is correct to write $D(%
\mathcal{O}_{\left\vert F\right\vert ^{2}}(iY))$ rather than $(D\mathcal{O}%
_{\left\vert F\right\vert ^{2}})(iY)$ as in (\cite{KOS}).)

Taking the adjoint of (\ref{d.form}), we obtain%
\begin{equation*}
D^{\ast }=C\left( \prod_{\alpha \in \Sigma ^{+}}\sin \left\langle \alpha
,Y\right\rangle \right) \left( \prod_{\alpha \in \Sigma ^{+}}D_{\alpha
}\right) .
\end{equation*}%
We now claim that%
\begin{equation}
\left( \prod_{\alpha \in \Sigma ^{+}}D_{\alpha }\right) e^{-\left\vert
Y\right\vert ^{2}/4t}=\left( \prod_{\alpha \in \Sigma ^{+}}\frac{%
-\left\langle \alpha ,Y\right\rangle }{2t}\right) e^{-\left\vert
Y\right\vert ^{2}/4t}.  \label{gaussder}
\end{equation}%
To see this, we first observe that the \textit{Fourier transform} of the
left-hand side of (\ref{gaussder}) is a constant times a Gaussian times the
polynomial $\pi .$ Since $\pi $ is alternating with respect to the action of
the Weyl group and since the Fourier transform commutes with the action of
the Weyl group, it follows that the left-hand side of (\ref{gaussder}) is
also alternating. The left-hand side of (\ref{gaussder}) is a polynomial $%
h(Y)$ times $e^{-\left\vert Y\right\vert ^{2}/4t},$ and the polynomial $h$
must be alternating. Furthermore, the leading order term in $h$ is easily
seen to be the polynomial appearing on the right-hand side of (\ref{gaussder}%
). The lower-order terms in $h$ are also alternating, and an alternating
polynomial whose degree is less than the number of positive roots must be
identically zero. (Compare Lemma 4 of \cite{U}.)

In the complex case, then, (\ref{fake.norm}) takes the form%
\begin{equation}
\left\Vert f\right\Vert ^{2}=C\frac{e^{t|\rho |^{2}}}{t^{d/2}}\int_{%
\mathfrak{a}^{+}}\mathcal{O}_{\left\vert F\right\vert ^{2}}(iY)\left(
\prod_{\alpha \in \Sigma ^{+}}\frac{\sin \alpha (Y)}{\alpha (Y)}\right)
e^{-\left\vert Y\right\vert ^{2}/4t}\left( \prod_{\alpha \in \Sigma
^{+}}\alpha (Y)\right) ^{2}dY,  \label{complex.fake}
\end{equation}%
where we have rearranged the polynomial factors in a convenient way and
where $d=\dim (G/K).$ (In the complex case, $\dim (G/K)=\dim \mathfrak{a}%
+2\left\vert \Sigma ^{+}\right\vert .$) We claim that in this case, (\ref%
{complex.fake}) actually makes sense. Specifically, $\mathcal{O}_{\left\vert
F\right\vert ^{2}}(iY)$ may be computed by the Gutzmer formula (\ref{gutzmer}%
) and the explicit formula (\ref{spherical.form}) for the spherical
functions then indicates that sine factors on the right-hand side of (\ref%
{complex.fake}) cancel all the singularities in $\mathcal{O}_{\left\vert
F\right\vert ^{2}}.$

Meanwhile, in the complex case the density for generalized polar coordinates
(integration of Ad-$K$-invariant functions on $\mathfrak{p}$) is given by%
\begin{equation*}
\mu (Y)=C\left( \prod_{\alpha \in \Sigma ^{+}}\alpha (Y)\right) ^{2}.
\end{equation*}%
(This is Theorem I.5.17 of \cite{He2} in the case where each $m_{\alpha }$
is equal to 2.) Also, the product over $\Sigma ^{+}$ of $\sin \alpha
(Y)/\alpha (Y)$ is just the Jacobian factor $j^{\mathrm{c}}(Y)^{1/2}$ of (%
\ref{jc}). Thus, if we rewrite (\ref{complex.fake}) as a limit of integrals
over $\mathfrak{a}_{R}^{+}$ and use the equality of (\ref{rint}) and (\ref%
{rint2}) we see that (\ref{fake.norm}) becomes%
\begin{equation*}
\lim_{R\rightarrow \infty }C\frac{e^{t\left\vert \rho \right\vert ^{2}}}{%
t^{d/2}}\int_{x\in G/K}\int_{Y\in T_{x}^{2R}(G/K)}\left\vert F(\exp
_{x}(iY/2))\right\vert ^{2}j^{\mathrm{c}}(Y)^{1/2}e^{-\left\vert
Y\right\vert ^{2}/4t}~dY~dx.
\end{equation*}%
This is nothing but the isometry formula established in Theorem \ref%
{globalisom.thm}, disguised by the change of variable $Y\rightarrow Y/2.$

Presumably, this line of reasoning could be used to give a rigorous
reduction of our isometry formula to that of \cite{KOS}. However, some care
would have to be given to the boundary terms in the integration by parts.

\appendix{}

\section{The Gutzmer-type formula of Faraut\label{faraut.app}}

In this appendix, we discuss Faraut's Gutzmer-type formula, established in 
\cite{Far1} and then in a stronger form in \cite{Far2}. We are particularly
concerned with the conditions under which this formula can be applied. In 
\cite{Far1}, Faraut established the Gutzmer formula under the assumption
that the Fourier transform of the restriction of $F$ to $G/K$ has compact
support. We will show that this result can easily be extended to any $F$ of
the form $F=e^{t\Delta /2}f,$ with $f\in L^{2}(G/K),$ something we require
in the proof of the isometry formula. Meanwhile, in \cite{Far2}, Faraut
established the Gutzmer formula under the assumption that $F$ is
square-integrable over (a domain in) $\Xi $ with respect to a nice $G$%
-invariant measure. We require the result of \cite{Far2} in the proof of the
surjectivity theorem.

First, fix $f\in L^{2}(G/K)$ and let $F:=e^{t\Delta /2}f.$ If the Fourier
transform of $f$ is $\hat{f}$ (in the notation established in Section \ref%
{partial_gen.sec}), then the Fourier transform of $F$ is given by $\hat{F}%
(\xi )=\hat{f}(\xi )e^{-t(\left\vert \xi \right\vert ^{2}+\left\vert \rho
\right\vert ^{2})/2}.$ Let $F_{n}$ be the function whose Fourier transform
is given by%
\begin{equation}
\hat{F}_{n}(\xi )=\hat{f}(\xi )e^{-t(\left\vert \xi \right\vert
^{2}+\left\vert \rho \right\vert ^{2})/2}\chi _{n}(\xi ),  \label{fn.est}
\end{equation}%
where $\chi _{n}$ is the indicator function of the ball of radius $n$ in $%
\mathfrak{a}^{\ast }.$ Since the Fourier transform of $F_{n}$ has compact
support, the hypotheses of the Gutzmer formula in \cite{Far1} hold. Thus, $%
F_{n}$ has a holomorphic extension to $\Xi $ and the Gutzmer formula in (\ref%
{gutzmer}) holds.

Meanwhile, according to Lemma 2.1 and Remark 2.2 of \cite{KOS}, for each $Y$
in $\Omega ,$ there exists $C_{Y}$ such that 
\begin{equation*}
\phi _{\xi }(e^{iY})\leq C_{Y}e^{\left\vert \xi \right\vert \left\vert
Y\right\vert },
\end{equation*}%
for all $\xi \in \mathfrak{a},$ and $C_{Y}$ may be taken to be a locally
bounded function of $Y.$ Then, using the Gutzmer formula and (\ref{fn.est}),
we see that $F_{n}$ converges in $L^{2}$ on each $G$-orbit $G\cdot e^{iY}$,
and the $L^{2}$ convergence is locally uniform as a function of $Y.$ This
means that the $F_{n}$'s are converging in $L_{\mathrm{loc}}^{2}$, which
then implies that the limiting function $\Phi $ is holomorphic. Since also
the restriction of $\Phi $ to $G/K$ is the $L^{2}$ limit of the $F_{n}$'s,
namely, $F,$ we conclude that $\Phi =F.$ By the continuity of the $L^{2}$
norm, then, we conclude that the Gutzmer formula holds for $F.$

Meanwhile, the paper \cite{Far2} establishes the Gutzmer formula for
weighted Bergman spaces. This means that we assume $F$ is holomorphic on a $%
G $-invariant domain $\mathcal{D}\subset \Xi $ that contains $G/K$ and with
the property that the intersection of $\Gamma $ with each $T_{x}(G/K)$ is
convex. We then assume that $F$ is square-integrable over $\mathcal{D}$ with
respect to a $G$-invariant measure $p$ that has a positive density that is
locally bounded away from zero. We let $\mathcal{B}^{2}(\mathcal{D},p)$
denote (in Faraut's notation) the space of holomorphic functions on $%
\mathcal{D}$ that are square-integrable with respect to $p.$ Faraut proves
that if $F\in \mathcal{B}^{2}(\mathcal{D},p)$, then: (1) the restriction of $%
F$ to $G/K$ is square-integrable, (2) the restriction of $F$ to each $G$%
-orbit inside $\mathcal{D}$ is square-integrable, and (3) the Gutzmer
formula holds.

Let us elaborate briefly on one point that is used in the proof of this form
of the Gutzmer formula. The result is that given $F\in \mathcal{B}^{2}(%
\mathcal{D},p),$ there exists a sequence $F_{n}\in \mathcal{B}^{2}(\mathcal{D%
},p)$ converging to $F$ in the norm topology of $\mathcal{B}^{2}(\mathcal{D}%
,p)$ such that the Fourier transform of $\left. F_{n}\right\vert _{G/K}$ has
compact support. The argument for the existence of such a sequence is
implicit in \cite{Far2}, but we felt it might be helpful to spell it out
explicitly, since this is the key to extending the Gutzmer formula to
functions in $\mathcal{B}^{2}(\mathcal{D},p).$

Faraut shows (Proposition 3.1) that the restriction map $R:\mathcal{B}%
^{2}(\Gamma ,\alpha )\rightarrow L^{2}(G/K)$ is bounded and injective. It
follows that the adjoint map $R^{\ast }:L^{2}(G/K)\rightarrow \mathcal{B}%
^{2}(\mathcal{D},p)$ is bounded with dense image. Thus, since functions
whose Fourier transform has compact support are dense in $L^{2}(G/K),$ given 
$F\in \mathcal{B}^{2}(\mathcal{D},p),$ we can choose $g_{n}\in L^{2}(G/K)$
with compactly supported Fourier transform such that $R^{\ast
}g_{n}\rightarrow F$ in $\mathcal{B}^{2}(\mathcal{D},p).$ Meanwhile, the
operator $RR^{\ast }$ is a convolution operator on $L^{2}(G/K)$ (see p. 104
in \cite{Far2}), which preserves the space of functions with compactly
supported Fourier transform. Thus the Fourier transform of $RR^{\ast }g_{n}$
has compact support, which means that $F_{n}:=R^{\ast }g_{n}$ is the desired
sequence in $\mathcal{B}^{2}(\mathcal{D},p).$

In the surjectivity theorem, we wish to apply the Gutzmer formula in the
case $\mathcal{D}=T^{R_{0}}(G/K)$ and $p$ is the measure associated to the
density $\alpha (Y)=\nu _{2t}^{\mathrm{c}}(Y)j^{\mathrm{c}}(Y)$ as in
Proposition \ref{partial.prop}.

\end{document}